\documentclass[12pt,preprint]{aastex}
\usepackage{amsmath}
\shorttitle{Energy Spectra of Cosmic Ray Leptons}
\shortauthors{Stawarz et al.}

\begin{document}

\title{On the Energy Spectra of GeV/TeV Cosmic Ray Leptons}

\author{{\L}ukasz Stawarz\altaffilmark{1,\,2}, Vah\'e Petrosian\altaffilmark{1,\,3}, \& Roger D. Blandford\altaffilmark{1}}

\email{stawarz@slac.stanford.edu}

\altaffiltext{1}{Kavli Institute for Particle Astrophysics and Cosmology,
Stanford University, Stanford CA 94305, USA}
\altaffiltext{2}{Astronomical Observatory of the Jagiellonian University, ul.
Orla 171, 30-244 Krak\'ow, Poland}
\altaffiltext{3}{Departments of Physics and of Applied Physics, 
Stanford University, Stanford CA 94305, USA}

\def\beq{\begin{equation}}
\def\eeq{\end{equation}}
\def\bar{\begin{eqnarray}}
\def\ear{\end{eqnarray}}
\def\tr{\tau_{rad}\,}
\def\tc{\tau_{coul}\,}
\def\tb{\tau_{brem}\,}
\def\trT{\tau_{rad,\,T}\,}
\def\trKN{\tau_{rad,\,KN}\,}
\def\tesc{\tau_{esc}\,}
\def\tloss{\tau_{loss}\,}
\def\tac{\tau_{acc}\,}
\def\tpp{\tau_{pp}\,}
\def\epm{$e^\pm$\,\,}

\begin{abstract}
Recent observations of cosmic ray electrons from several instruments have revealed 
various degrees of deviation in the measured electron energy distribution from a simple power-law, 
in a form of an excess around $0.1$ to $1$\,TeV energies. An even more prominent
deviation and excess has been observed in the fraction of cosmic ray positrons around
$10$ and $100$\,GeV energies. These observations have received considerable attention and many
theoretical models have been proposed to explain them. The models rely on either
dark matter annihilation/decay or specific nearby astrophysical sources, and involve
several additional assumptions regarding the dark matter distribution or particle 
acceleration. In this paper we show that the observed excesses in the electron spectrum 
may be easily re-produced without invoking any unusual sources other than the general diffuse 
Galactic components of cosmic rays. The model presented here assumes a power-law 
injection of electrons (and protons) by supernova remnants, and evaluates their expected 
energy spectrum based on a simple kinetic equation describing the propagation of 
charged particles in the interstellar medium. The primary physical effect involved 
is the Klein-Nishina suppression of the electron cooling rate around TeV energies. 
With a very reasonable choice of the model parameters 
characterizing the local interstellar medium, we can reproduce the most recent 
observations by Fermi and HESS experiments. Interestingly, in our model the injection 
spectral index of cosmic ray electrons becomes comparable to, or even equal to that 
of cosmic ray protons. The Klein-Nishina effect may also affect the propagation of 
the secondary \epm pairs, and therefore modify the cosmic ray positron-to-electron ratio. 
We have explored this possibility by considering two mechanisms 
for production of \epm pairs within the Galaxy. The first is due to the decay of 
$\pi^\pm$'s produced by interaction of cosmic ray nuclei with ambient protons. 
The second source discussed here is due to the annihilation of the diffuse Galactic 
$\gamma$-rays on the stellar photon field. We find that high positron 
fraction increasing with energy, as claimed by the PAMELA experiment, cannot be explained 
in our model with the conservative set of the model parameters. We are able, however, 
to reproduce the PAMELA (as well as Fermi and HESS) results assuming high values of the 
starlight and interstellar gas densities, which would be more appropriate for vicinities 
of supernova remnants. A possible solution to this problem may be that cosmic rays
undergo most of their interactions near their sources due to the efficient trapping 
in the far upstream of supernova shocks by self-generated, cosmic ray-driven turbulence.
\end{abstract}

\keywords{cosmic rays --- Galaxy: general --- ISM: general}

\section{Introduction}

Measurements of the energy spectra of cosmic ray (CR) species are of a great
importance for understanding the physics of Galactic CR sources (such as pulsars
or supernova remnants), as well as for constraining the internal structure of
Milky Way, since this structure (topology and intensity of the Galactic magnetic
field, profiles and energy distribution of different Galactic photon fields,
distribution of interstellar gas and dust, etc.) determines the spatial and
energy evolution of the injected charged particles which propagate through the
interstellar medium. In addition, as discussed in several papers by a number of
authors \citep[e.g.,][]{jun96,che02}, annihilation or decay of a dark matter
(Kaluza-Klein particles or supersymmetric WIMPs) may also imprint some
signatures in the observed spectra of CR electrons within the GeV--TeV energy
range. Consequently,  the most recent observations by the ATIC, PAMELA, Fermi,
and HESS experiments \citep[][respectively]{cha08,adr09,abd09,aha08,aha09a} have
generated significant interest on this topic in astrophysic and particle physics
communities.

The most sophisticated framework for analyzing the CR propagation within the
Galaxy is provided by the GALPROP model presented first by \citet{mos98} and
\citet{str98}. This model assumes injection of a power-law electrons and nuclei,
and follows their spatial and energy evolution under the influence of different
radiative processes (Coulomb losses, synchrotron and IC cooling, proton-proton 
collisions, etc.), taking also into
account the relevant interactions of charged particles with the interstellar
turbulent magnetic field (with the assumed Kolmogorov spectrum) in a quasi-linear
approximation regime. The model can successfully explain many findings
regarding the hadronic CR spectrum and its composition \citep{mos02,mos03}, as
well as the observed Galactic diffuse $\gamma$-ray emission \citep{mos00,str00,str04,por08}. 
However, the model predicts also the decrease of the CR positron-to-electron 
ratio with particles' energy in the GeV--TeV range, in a disagreement with the 
observational indications. This prediction was made under the working hypothesis 
that bulk of the Galactic \epm pairs are created in the collisions of relativistic 
CR protons with ambient gas, and the subsequent decays of the generated pions
\citep[see the discussion in][]{mos98}. One should note that the previous (prior 
to 2008) measurements regarding this issue were restricted to electron energies 
$E_e <10$\,GeV, and as such could be seriously affected by the charge dependence 
of solar modulation. On the other hand, the most recent PAMELA observations, 
reaching $E_e \simeq 100$\,GeV energies, confirmed the increasing positron 
fraction in the CR spectrum \citep{adr09}.

The other challenge to the `standard' model of CR propagation came from the
observations of ATIC collaboration, which reported a sharp pile-up around $E_e
\simeq 0.5$\,TeV above the power-law spectrum both observed at lower energies 
but also emerging from the GALPROP calculations \citep{cha08}. Reality of
this sharp spectral feature has been questioned by the Fermi and HESS experiments 
\citep[see][]{abd09,aha09a}, which show a much smaller and broader excess over the 
best fit power-law continuum $J_e(E_e) \propto E_e^{-3}$. In this context one has 
to keep in mind that due to rapid radiative losses of the TeV-energy electrons, 
their spectrum measured in the Solar System may be possibly dominated by a few 
local sources \citep[most likely nearby pulsars such as Vela or Geminga; see,
e.g.,][]{she70,nis80,aha95,kob04}, and therefore may be more complex than a
featureless power-law continuum. In particular, it may reflect the
non-stationary and stochastic nature of such sources \citep{poh98,gra09}. 
More interestingly, the substructure  around $0.5$\,TeV observed by ATIC was 
argued to be consistent with that  expected from the annihilation of the 
Kaluza-Klein dark matter particles \citep{cha08}. Both `local pulsar' and 
dark matter scenarios were claimed to successfully account for the increasing 
positron fraction in the CR spectrum as well \citep[e.g.,][]{poh09,gra09}.

Obviously, the dark matter interpretation of the ATIC, Fermi, HESS and PAMELA results 
is of a great interest. However, in order to fit the collected data in the framework
of this model, large though arbitrary `boost factors' have to be invoked for
the dark matter annihilation/decay fluxes, which are only roughly justified by a
possible non-uniform (clumpy) distribution of the dark matter near the Solar
System \citep[e.g.][]{cha08,ela09,hoo09}. In addition, as shown by \citet{pro09}, 
the dark matter scenario for the positron excess conflicts with the observations 
of the extragalactic background radiation in the X-ray/$\gamma$-ray energy range 
\citep[see also][]{bel09}. Similarly, the `local pulsar' interpretation also relies 
on several model assumptions (regarding particle acceleration in relativistic 
outflows), which are not observationally verified yet.

In this paper we explore the possibility of explaining the aforementioned data 
sets by a simple model for the generation and propagation of CRs in the Galaxy
(and the vicinity of Earth), without invoking any new/unconventional sources of
the TeV-energy electrons or positrons. This scenario is not intended to compete
with the complexity of the GALPROP model. It is intended instead to point out
several effects which, even though being standard and at some level inevitable,
might have been underestimated or overlooked in the previous analysis. The new
physics involved here is that  we emphsize  the importance of the Klein-Nishina 
(KN) suppression of the inverse Compton (IC) scattering cross section for
ultrarelativistic electrons and positrons interacting with the Galactic
starlight and other photon fields. We evaluate the modification of the primary
electron energy spectrum which results from this new aspect, as well as that of
secondary pairs arising from the annihilation of the Galactic $\gamma$-rays on
starlight and from CR proton interactions with the interstellar medium (ISM). 
In \S\,2 we discuss our model and present the kinetic equation 
describing the propagation of CR electrons (as well as protons and positrons). 
We first describe the KN effect qualitatively and then compare the CR electron 
spectra obtained from detailed solution of the kinetic equation with observations. 
In \S\,3 we consider various mechanisms for production of secondary \epm pairs 
and derive their spectra using the same propagation model, comparing again the 
resulting spectra with those observed by PAMELA. A brief discussion and summary 
is presented in \S\,4.

\section{Primary CR Electrons}

We start this section with the description of the relevant interactions of CR
electrons (accelerated and injected into the Galaxy by, presumably, supernova remnants)
with the ISM photons (mainly starlight), magnetic field, and turbulence. We then
describe the resultant spectra of these `primary' electrons and compare them with 
the observations reported by ATIC, Fermi, HESS, and other experiments.

\subsection{Interactions of CR Electrons}

The propagation of relativistic electrons injected into the ISM is determined by
two basic interactions: radiative cooling and interactions with plasma turbulence. 
The latter causes diffusion in space (determining the rate of the escape of electrons 
from the Galaxy) and diffusion in energy (determining the rate of the acceleration).
For the electrons with energies above GeV, the radiative cooling is mainly via the 
IC scattering on ambient photon fields and due to the synchrotron emission in the 
Galactic magnetic field. The relevant photon 
fields are the Cosmic Microwave Background (CMB) radiation with the energy density 
$u_{cmb} \simeq 0.26$\,eV\,cm$^{-3}$, the Galactic starlight, and far infrared 
photons from the dust emission. The latter two are expected to dominate over 
the CMB (i.e., $u_{star},\,u_{dust} > u_{cmb}$) in the inner parts of the 
Galactic disk ($r \lesssim 10$\,kpc from the center) by a factor of at least a few 
\citep[see, e.g.,][]{str00,por06,mos06,por08}. For an electron with energy $E_e$, the
characteristic cooling timescale in the Thomson (T) regime is therefore given as
\bar
& & \trT \simeq {3 \, m_e^2 c^3 \over 4 \sigma_T \,u_{tot}\,E_e} \, , \quad 
{\rm with} \nonumber \\ & & u_{tot} \equiv u_{cmb}\,(1+ \xi) +  {B^2 \over 8 \pi} 
\quad {\rm and} \quad \xi \equiv {u_{dust}+u_{star} \over u_{cmb}} \, .
\ear
From this one gets $\trT \simeq 10^9\,(E_e/{\rm GeV})^{-1}\,(1 + 0.1\,B_{\rm \mu
G}^2+\xi)^{-1}$\,yr, which, for the illustrative (though expected) Galactic
magnetic field $B_{\rm \mu G} \equiv B/{\rm \mu G} \simeq 3$ and $\xi \simeq
10$, leads to $\trT \simeq 100\,(E_e/{\rm GeV})^{-1}$\,Myr. This estimate breaks
down at high particle (and/or photon) energies where the KN effect reduces the
IC cross section. This happens when the target photon energy in the electron
rest frame exceeds electron rest mass, which, in the observer frame, translates 
to $\varepsilon > m_e^2 c^4 / 4 E_e \simeq 65\,(E_e/{\rm GeV})^{-1}$\,eV \citep{blu70}.

For the evaluation of the interactions of CR electrons with magnetic turbulence we
use the quasi-linear approximation for the particle-wave interactions \cite[see,
e.g.,][and references therein]{sch02} and assume a Kolmogorov spectrum
of turbulence as a superposition of magnetohydrodynamical (MHD) waves. The
corresponding spatial diffusion timescale may be estimated as $\tesc \simeq
3\,\ell^2/c\,\lambda\!(E_e)$, where $\ell$ is the linear scale of the system
and the particle mean free path $\lambda(E_e) \simeq \zeta \, r_g^{1/3} \,\lambda_{max}^{2/3}$. 
Here $\zeta \equiv (B/\delta B)^2$ is the ratio of energy densities stored in 
the large-scale (`unperturbed') and turbulent magnetic fields, $r_g \equiv E_e 
/ e B \simeq 3 \times 10^{12}\,(E_e/{\rm GeV})\,B_{\rm \mu G}^{-1}$\,cm is 
the electron gyroradius, and $\lambda_{max}$ is the maximum wavelength of the 
turbulent modes. This gives $\tesc \simeq 10^7\,\zeta^{-1}\,B_{\rm \mu G}^{1/3}
\,(\ell/{\rm kpc})^2\,(\lambda_{max}/{\rm kpc})^{-2/3}\,(E_e/{\rm GeV})^{-1/3}$\,yr 
which, for the ISM parameters $B_{\rm \mu G} \simeq 3$, $\zeta \simeq 1$, and 
$\lambda_{max} \simeq 1$\,kpc, simplifies further to $\tesc \simeq 10 \, 
(\ell/{\rm kpc})^2\,(E_e/{\rm GeV})^{-1/3}$\,Myr. In other words, an electron 
with energy $E_e$ travels the distance $\ell \simeq 3\,(E_e/{\rm GeV})^{-1/3}$\,kpc 
within ISM before loosing its energy via radiative cooling (for $\xi \simeq 10$). 
Hence, high-energy CR electrons ($E_e > 10$\,GeV) detected near the Earth 
are supposed to originate from local ($\ell < 3$\,kpc) region and recently 
operating ($t < 100$\,Myr) sources \citep{she70}.

It should be noted in this context that several observational findings
(regarding, e.g., secondary-to-primary ratios of some CR elements) are better 
interpreted in terms of the particle diffusion shaped by the ISM turbulence 
characterized by the Kraichnan energy spectrum, rather than of the Kolmogorov 
form anticipated above \citep[see][and references therein]{ptu06,str07}. 
If this is the case indeed, then 
$\lambda(E_e) \simeq \zeta \, r_g^{1/2} \,\lambda_{max}^{1/2}$ and 
$\tesc \simeq 500 \, (\ell/{\rm kpc})^2\,(E_e/{\rm GeV})^{-1/2}$\,Myr, and
hence the conclusion regarding the local origin of $> 10$\,GeV energy
CR electrons is even strengthened.

On the other hand, the order-of-magnitude estimate presented above should be taken
with caution, because of the uncertainty in the value of $\lambda_{max}$, 
which is not a directly measured quantity, but is only expected to be roughly 
of the scale of the Galactic disk thickness, $\sim 1$\,kpc. In addition, 
the use of the Kolmogorov/Kraichnan spectrum for ISM turbulence is only partly justified 
by observations/theoretical models \citep[see, e.g.,][and references
therein]{cho03,str07}. Finally, the anticipated quasi-linear approximation is
appropriate for modeling of the particle diffusion \emph{along} the magnetic 
field lines (and thus rather \emph{within} the Galactic disk), and is not
expected to describe properly the propagation of particles \emph{accross} the 
magnetic field lines (i.e., the escape of particles from the Galactic disk).

The spatial diffusion of CR electrons is accompanied by their diffusion in the
momentum space, leading to a net acceleration of particles on the characteristic
timescale $\tac \simeq 3\,\lambda\!(E_e) / c \beta_{sc}^2$, where $\beta_{sc}$
is the velocity of turbulent MHD modes in the units of speed of light
\citep[e.g.,][]{bla87}. For a low-beta plasma, the MHD wave velocity is equal 
to the Alfv\'en velocity $c \beta_{sc} \simeq v_{A} = B/(4\pi m_p n_{ism})^{1/2} 
\simeq 2 \times 10^5 \, B_{\rm \mu G}\,(n_{ism}/{\rm cm^{-3}})^{-1/2}$\,cm\,s$^{-1}$, 
where $n_{ism}$ is the number density of the ambient plasma. Thus, for a Kolmogorov 
spectrum of the ISM turbulence $\tac \simeq 10^{11}\,\zeta\,\,B_{\rm \mu G}^{-7/3}\,
(\lambda_{max}/{\rm kpc})^{2/3}\,(n_{ism}/{\rm cm^{-3}})\,(E_e/{\rm GeV})^{1/3}$\,yr. 
Assuming again values $B_{\rm \mu G} \simeq 3$, $\zeta \simeq 1$, $\lambda_{max} 
\simeq 1$\,kpc, and $n_{ism} \simeq 1$\,cm$^{-3}$, this simplifies further to 
$\tac \simeq 10^4\,(E_e/{\rm GeV})^{1/3}$\,Myr, which indicates that ultrarelativistic 
electrons undergo little turbulent acceleration within ISM before they cool 
radiatively or escape \citep[see in this context, e.g.,][]{ptu06}. Thus, radiative 
cooling and diffusive escape are the main processes controlling evolution of CR 
electrons. Again, in the case of the Kraichnan form of the magnetic turbulence, 
the same ISM parameters give $\tac \simeq 400\,(E_e/{\rm GeV})^{1/2}$\,Myr, which
is still much longer than the radiative cooling timescale of electrons with 
$E_e > 10$\,GeV.

\subsection{Transport Equation}

In what follows we assume that CR electrons are injected by numerous sources
throughout the ISM which has a smooth and slowly varying distribution of gas,
photons, and magnetic fields relative to the relevant interaction scales
discussed above. As indicated by the small mean free path calculated above, 
the electrons undergo multiple scattering before any other interactions so 
that they acquire an isotropic pitch angle distribution. As a result, for the 
relevant scales of about few kpc in the vicinity of the Earth we can use the 
homogeneous and isotropic approximation in describing the transport of the CR 
electrons, where the spatial diffusion can be represented by an overall 
(energy dependent) escape term. Hence, ignoring turbulent acceleration, the 
propagation of the GeV/TeV CR electrons within the local ISM can be described 
by the following kinetic equation:
\beq\label{keq}
{\partial n_e\!(E_e) \over \partial t} = - {\partial \over \partial E_e}
\left[{E_e\, n_e\!(E_e) \over \tloss}\right]  - {n_e\!(E_e) \over \tesc} + 
{\dot Q}_e\!(E_e) \,
\eeq
\citep[see, e.g.,][and references therein]{pet04}, where ${\dot Q}_e(E_e)$ denotes 
the injection rate of electrons, and $\tloss$ the total energy losses timescale. 
The steady-state solution to this equation reads as
\beq\label{solution}
n_e\!(E_e) = {\tloss\!(E_e) \over E_e} \, \int_{E_e}^{\infty} dE_e' \,\, 
{\dot Q}_e\!(E_e') \,\, \exp\!\left[- \int_{E_e}^{E_e'} {dE_e'' \over E_e''} 
\,\, {\tloss\!(E_e'') \over \tesc\!(E_e'')} \right] \, .
\eeq
For a broad, e.g. a power-law-type injection function, a very rough but
illustrative approximations to this solutions are $n_e(E_e< E_0) \sim \tesc 
\times {\dot Q}_e$ and $n_e(E_e > E_0) \sim \tloss \times {\dot Q}_e$, where at
$E_0$ we have $\tloss(E_0) = \tesc(E_0)$. Note that for $\tloss \simeq \trT$ and
$\tesc$ as specified above ($\xi \simeq 10$, and $\ell \simeq 3$\,kpc, roughly the
vertical scale of the Galactic disk) one has $E_0 \simeq 1$\,GeV. Thus, from the
precisely known transport timescales and the observed electron flux $J_e(E_e >
{\rm GeV}) \propto E_e^{-3}$ one can set direct constraints for the electron
injection function ${\dot Q}_e(E_e)$. We note in this context that 
non-relativistic shock waves associated with Galactic supernova remnants 
(SNRs) are expected to be the primary source of the observed high-energy 
($>10$\,GeV) CRs \citep[e.g.,][]{bla78}. This anticipation seems 
to be confirmed by the most recent X-ray and $\gamma$-ray observations regarding 
several remnants \citep[see, e.g.,][and references therein]{vin08}. However, 
the injection spectrum of CR particles (both leptons and hadrons) may be 
substantially different from the energy spectra of freshly accelerated electrons
and ions at SNR shocks, due to a complex convolution of particle cooling, transport, 
and CR-driven magnetic amplification processes in vicinity of non-linear shock waves
\citep{bla87,cap09}.

In this paper we restrict the analysis to the CR electrons not affected by the
solar modulation, i.e. the ones with energies $E_e > 10$\,GeV, for which the escape effects
can be neglected. Therefore we omit the escape term in the equation~\ref{keq}, and obtain
\beq\label{simpkeq}
n_e(E_e) = {\tloss\!(E_e) \over E_e} \, \int_{E_e}^{\infty} dE_e' \,\, {\dot Q}_e\!(E_e') \, ,
\eeq
so that for the injected function ${\dot Q}_e(E_e) \propto E_e^{-s_e}$ and $\tloss
\simeq \trT \propto E_e^{-1}$ the observed electron spectrum will also be a power-law, 
$n_e(E_e) \propto E_e^{-s_e-1}$. Hence $s_e = 2$ will be required by the observed electron 
flux $J_e(E_e) \propto E_e^{-3}$. However, if the dominant electron cooling is due to the 
IC scattering in the KN regime, the energy losses time scale goes roughly as $\tloss 
\simeq \trKN \propto E_e^{1/2}$, which will give rise to $n_e(E_e) \propto E_e^{-s_e+0.5}$. 
In other words, \emph{within the energy range where the IC/KN losses are dominant}, the
steady-state electron energy distribution is expected to pile-up above that expected 
from extrapolation of the Thomson-regime spectrum, as discussed in
different contexts by, e.g., \citet{aha85,der02,kus05,mod05}, and \citet{sta06}. 
These KN-related spectral pile-ups are more and more pronounced for flatter and 
flatter injection continuum\footnote{Note that
the pile-up effects in the electron energy distribution resulting from the KN
suppression are present at some level for any slope of the injection spectrum,
unlike the analogous effects related to the synchrotron cooling alone \citep{kar62}, 
which are present only for $s_e < 2$.}. Interestingly, as already noted in the 
previous section, for the characteristic energy of the starlight photons 
$\varepsilon_{star} \simeq 1$\,eV (wavelengths $\lambda_{star} \simeq 1$\,$\mu$m), 
this is expected to happen for $E_e \gtrsim m_e^2 c^4 / 4 \varepsilon_{star} 
\sim 0.1$\,TeV, i.e. within the range of the claimed `electron excess', assuming 
that the energy density of the starlight emission dominates all the other Galactic 
photon and magnetic fields ($\xi > 1$). As a result, if the discussed KN effect
plays a role, a relatively steep electron injection index $s_e > 2$ is in fact 
required to account for the observed spectrum $J_e(E_e) \propto E_e^{-3}$. 

\subsection{Energy Spectra of CR Electrons}

\begin{figure}
\epsscale{0.45}
\plotone{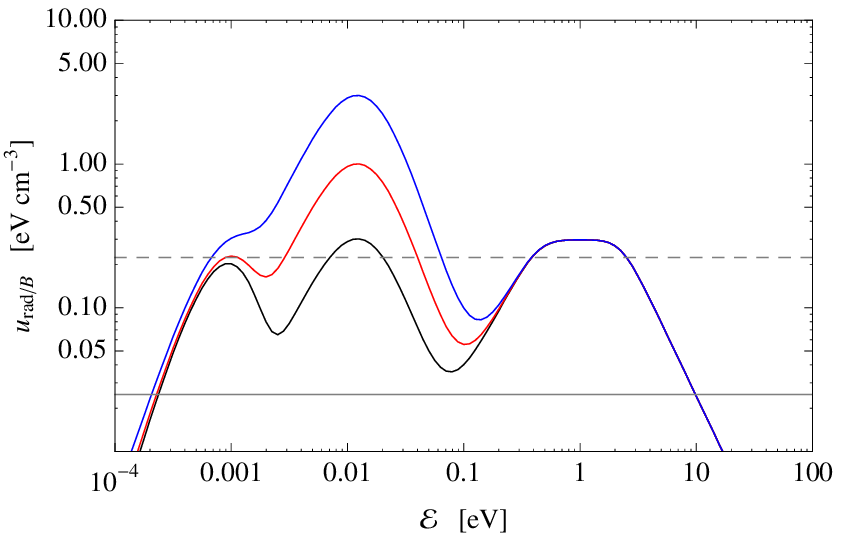}
\plotone{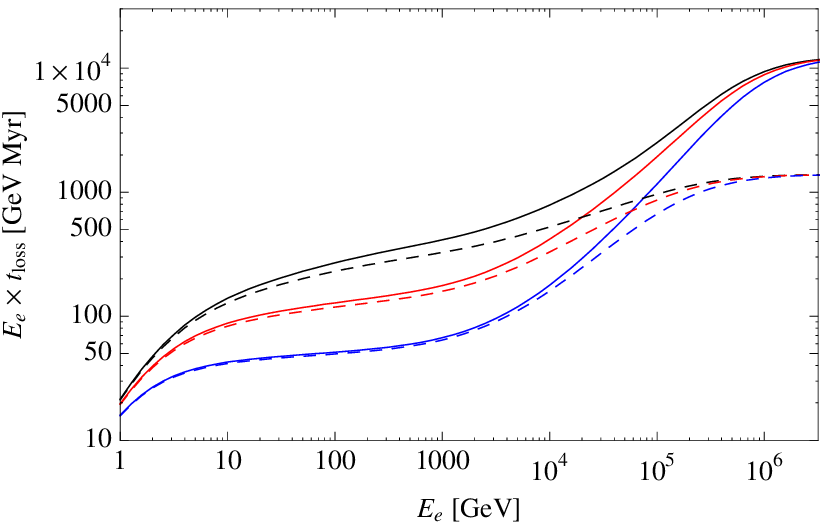}
\plotone{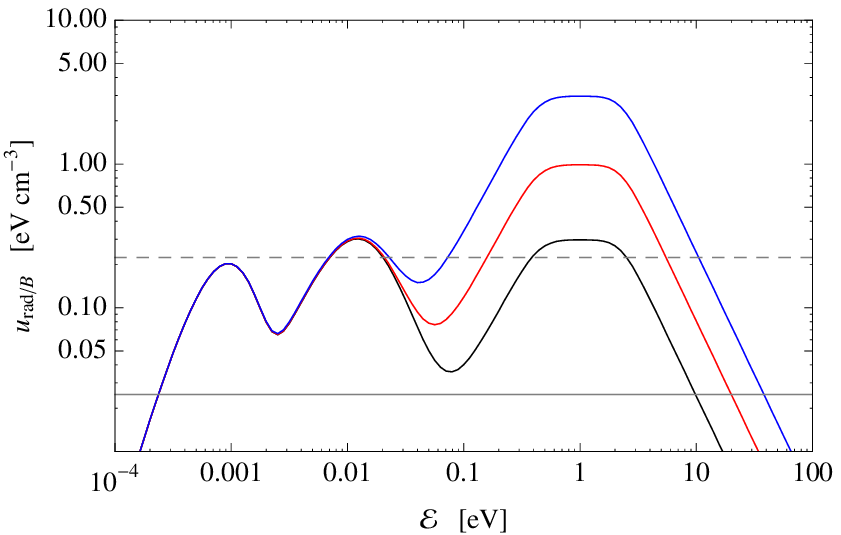}
\plotone{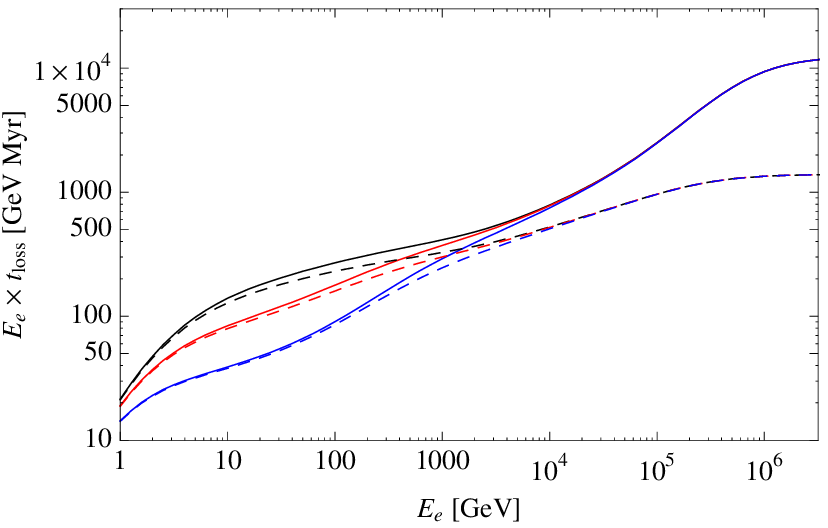}
\plotone{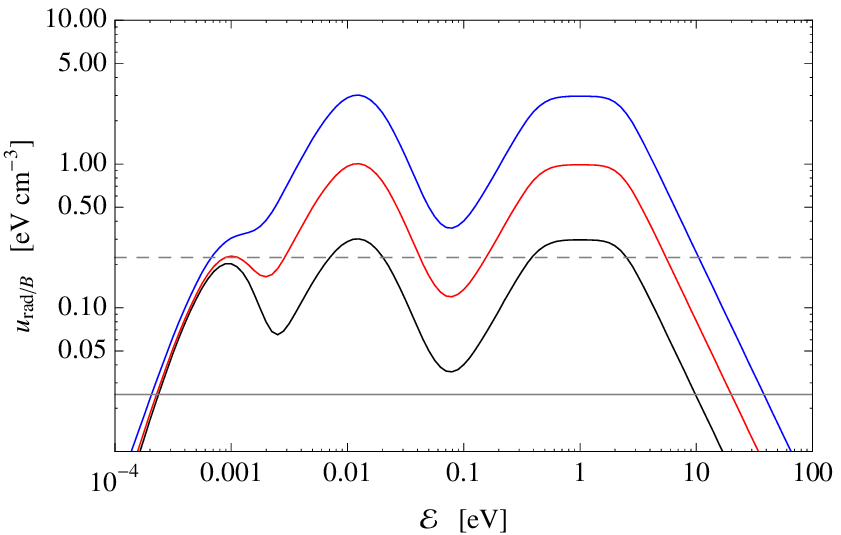}
\plotone{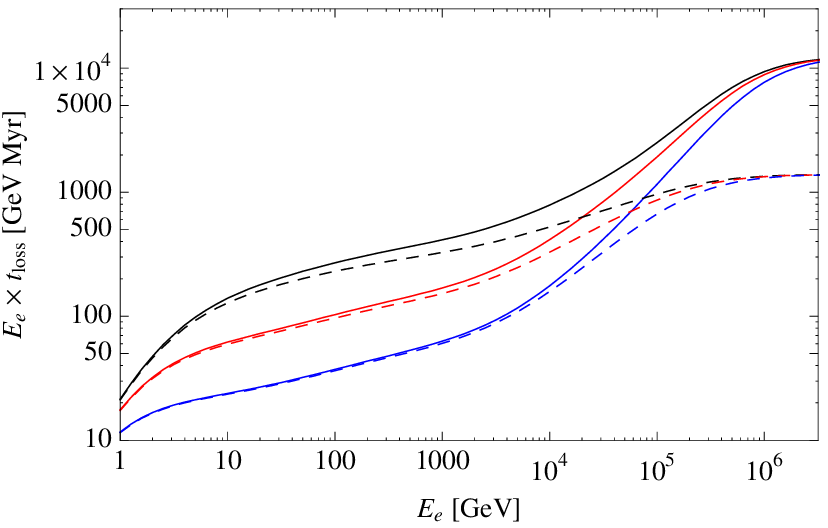}
\caption{
{\bf Left panels:} Models of the target photon fields including the CMB and
different valuse of starlight and dust emission. Each panel has three curves for
$u_{dust}$ and/or $u_{star}$ equal to $0.3$, $1$, or $3$\,eV\,cm$^{-3}$,
varying independently or together (black, red, and blue curves). In the top panel
$u_{star}$ is set as $0.3$\,eV\,cm$^{-3}$. In the middle panel $u_{dust}$ is set 
as $0.3$\,eV\,cm$^{-3}$. In the bottom panel $u_{star} = u_{dust}$.
Two different values of the magnetic field densities are also shown: 
$B = 1$\,$\mu$G (solid horizontal lines), and $3$\,$\mu$G
(dashed horizontal lines). {\bf Right panels:} The energy dependence of the 
energy losses timescales (multiplied by energy) for the Galactic CR electrons
corresponding to the different levels of the Galactic photon and magnetic fields
shown on the left panels, and $n_{ism} = 1$\,cm$^{-3}$.}
\label{lossrate}
\end{figure}

We now present a more rigorous treatment of the KN effect and show that it may be 
the primary cause of the observed deviation of CR electron spectrum from a simple
power law. For this purpose we need a more detailed description of the Galactic
photon fields and of the radiative cooling rate. We model the Galactic dust and
starlight emission by the functions $u(\varepsilon)$ similar to the ones given by
\citet{por06}, which peak for $\varepsilon \simeq 0.015$\,eV and $\varepsilon
\simeq 0.5-2.0$\,eV photon energies with the maximum levels $u_{dust}$ and
$u_{star}$, respectively, such that the total (integrated over $\varepsilon$)
energy densities in these components are $\simeq 2 \, u_{dust}$ and $\simeq 3 \, 
u_{star}$. The left panels of Figure\,\ref{lossrate} show variation with photon 
energy of the CMB, dust and starlight energy densities for several relative values 
of these densities. In each panel we show three curves: for three values of
$u_{dust}$ and fixed $u_{star}=0.3$\,eV\,cm$^{-3}$ (top); three values of
$u_{star}$ and $u_{dust}=0.3$\,eV\,cm$^{-3}$ (middle); three values of 
$u_{dust}=u_{star}$ (bottom). We also show the energy density of the magnetic 
field for $B = 1$\,$\mu$G (solid horizontal lines), and $3$\,$\mu$G (dashed 
horizontal lines).

Follwing \citet{mod05}, we approximate the radiative energy loss timescale for
ultrarelativistic leptons as
\beq
\tr\!(E_e) \simeq {3 m_e^2 c^3 \over 4 \sigma_T E_e} \, \left[ {B^2 \over 8 \pi} + \int
d\varepsilon \, u_{tot}\!(\varepsilon) \,\, f_{KN}\!\!\left({4 E_e \,
\varepsilon \over m_e^2 c^4}\right)\right]^{-1} \, ,
\eeq
where  the total energy density of the Galactic photon fields (including CMB
radiation) is $\int d\varepsilon \, u_{tot}(\varepsilon)$. Here the KN
correction factor is taken to be of the form
\beq
f_{KN}(x) \simeq \left\{ \begin{array}{ccc} (1+x)^{-1.5} & {\rm for} & x < 10^4
\\
{27 \over 2}\,x^{-2}\,\left(\ln x - {11 \over 6}\right) & {\rm for} & x > 10^4
\end{array} \right. \, .
\eeq
In addition, we consider the electron energy losses due to the Coulomb collisions
and the bremsstrahlung process. The appropriate timescale of these can be approximated 
as, respectively,
\beq
\tc\!(E_e) \simeq {2 E_e \over 3 \ln\!\Lambda \, m_e c^3 \sigma_T \, n_{ism}} \, ,
\eeq
where $\ln \Lambda \simeq 40$ \citep{pet73}, and
\beq
\tb\!(E_e) \simeq {2 \pi \over 3 \alpha_{fs} \sigma_T c \, n_{ism} \, \chi\!(E_e)} \, ,
\eeq
where $\alpha_{fs} \simeq 1/ 137$ is the fine structure constant and $\chi\!(E_e) \simeq \ln\left(2 E_e 
/ m_e c^2\right) - 1/3 \simeq 10$ \citep{pet01}. We note that the above form of $\tb$ includes 
electron-ion and electron-electron bremsstrahlung assuming completely unscreened limit with 
$10\%$ fully ionized helium abundance. With such, the total energy losses timescale
is
\beq
\tau_{loss}^{-1}(E_e) = \tau_{coul}^{-1}(E_e) + \tau_{brem}^{-1}(E_e) + 
\tau_{rad}^{-1}(E_e) \, .
\eeq

The right panels of Figure\,\ref{lossrate} show the total energy losses timescales
(multiplied by energy) corresponding to the photon and $B$ field energy densities
the same as in the left panels, and $n_{ism} = 1$\,cm$^{-3}$. As evident, 
at low energies $E_e \lesssim 1$\,GeV the Coulomb and bremsstrahlung 
processes dominate electron cooling, since $\tc \sim 50\,(E_e/{\rm GeV})$\,Myr and 
$\tb \sim 50$\,Myr (for $n_{ism} = 1$\,cm$^{-3}$ and $\chi\!(E_e) = 10$).
At higher electron energies, the IC/T losses take over. However, for $E_e > 10$\,GeV
the radiative cooling rates deviate from the ones characterizing the Thomson
regime (which would be represented by horizontal lines on these plots) due to
the KN effect. Note that these deviations are the strongest in the case of a
large ratio $u_{star}/u_{dust}$. This is because for dust emission in the
far-infrared range, the dominant radiative cooling is still in the Thomson
regime even for relatively energetic electrons. Therefore, the KN suppression
for the optical target photons becomes important only for large values of the
ratio $u_{star}/u_{dust}$. \emph{This is exactly the reason why the KN-related 
features in the CR electron spectrum discussed here may remain unnoticed in the 
GALPROP calculations, even though this code includes the exact prescription of 
the IC cross section, valid in both T and KN regimes.}

Using this radiative loss rate in  the  simplified version of the kinetic
equation (\ref{simpkeq}) we obtain the energy flux spectrum $J_e(E_e) 
\propto n_e(E_e)$ of CR electrons. For the injection function ${\dot Q}_e(E_e)$ 
we use
\beq
{\dot Q}_e(E_e) = k_e \, E_e^{-s_e} \, \exp\left[ - {E_e \over E_{e,\,max}}\right] \, ,
\eeq
with the normalization $k_e$ fixed so that $[E_e^3 J_e(E_e)]_{E_e =
30\,{\rm GeV}} = 151.4$\,GeV$^{2}$\,m$^{-2}$\,s$^{-1}$\,sr$^{-1}$, as indicated
by the Fermi data. In general, the model outlined above has seven free parameters, 
namely $s_e$, $E_{e,\,max}$, $u_{dust}$, $u_{star}$, $B$, $n_{ism}$, and $\ell$. 
However, we fix for illustration $\ell = 3$\,kpc, $n_{ism} = 1$\,cm$^{-3}$, and 
$E_{e,\,max} = 2$\,TeV, so that we are left with only four free parameters 
$s_e$, $u_{dust}$, $u_{star}$, and $B$. We note in this context that the 
direct measurements of the Galactic photon and magnetic field energy densities 
are difficult due to substantial foregrounds, and thus the associated uncertainties 
are relatively large \citep[e.g.,][]{cru03,hau01}. Below we explore the 
corresponding parameter space of the model.

\begin{figure}
\epsscale{0.45}
\plotone{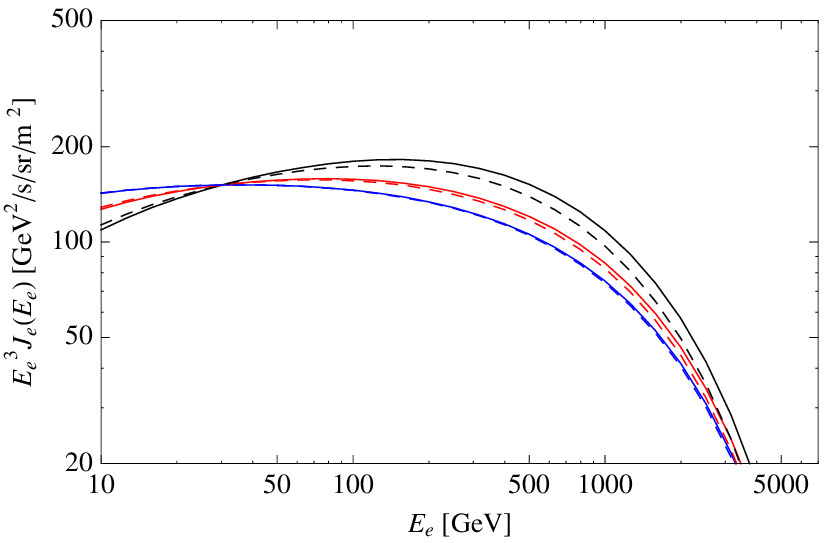}
\plotone{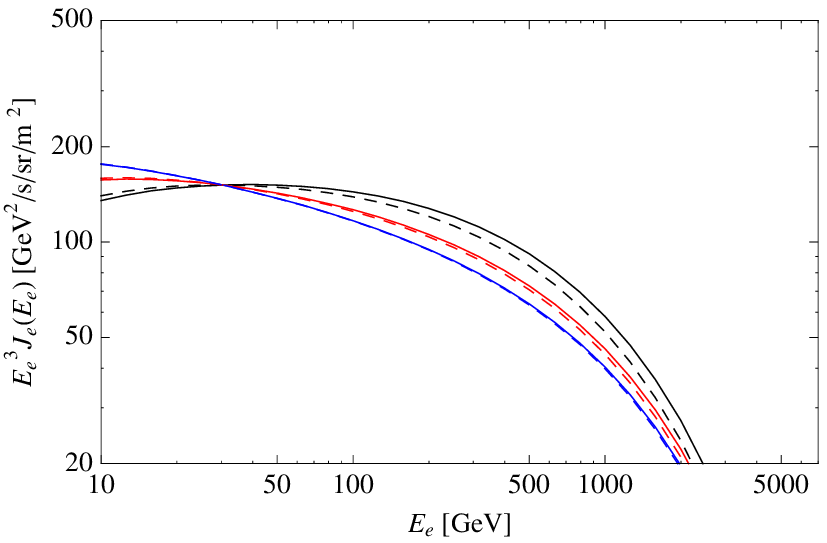}
\plotone{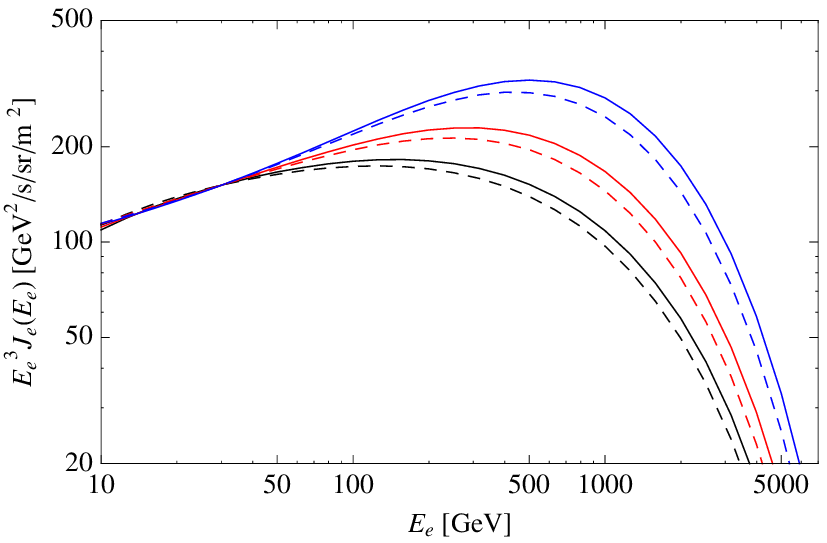}
\plotone{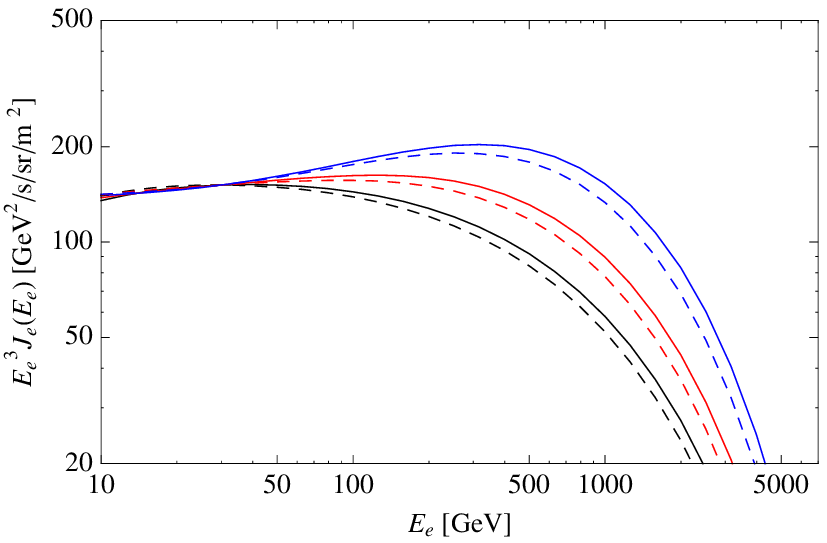}
\plotone{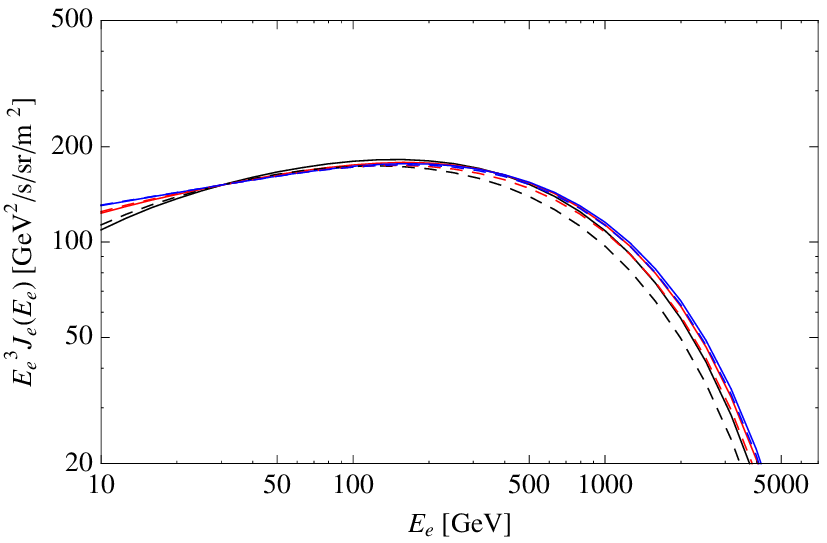}
\plotone{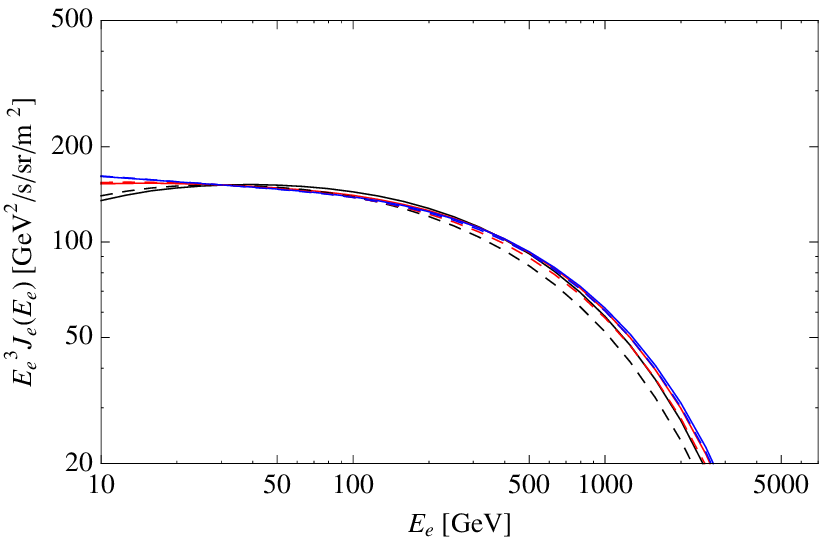}
\caption{The energy spectra of primary electrons corresponding to two different
injection spectral indices, $s_e = 2.0$ ({\bf left panels}) and  $2.2$ ({\bf right panels}), 
for the same set of the model parameters given in Figure\,\ref{lossrate}}
\label{especs}
\end{figure}

Figure\,\ref{especs} shows the energy spectra of primary electrons corresponding
to two different injection spectral indices, $s_e = 2.0$ and $2.2$, and to the
same choice of the values of the other three model parameters used
in Figure\,\ref{lossrate}. As evident, the expected KN pile-up effects are indeed
present, being the most pronounced for flatter injection continuum, and for large
values of the ratio $u_{star}/u_{dust}$. The value for the ISM magnetic field
have little effect on the results, as long as $B < 10$\,$\mu$G. One conclusion
here is that different combinations of the parameters $s_e$, $u_{star}$, and
$u_{dust}$ can lead to the observed electron spectrum $J_e(E_e) \propto
E_e^{-3}$. However, our primary result is that it is relatively easy to account 
for a possible minor excess in the energy distribution of primary CR electrons 
over this power law in the $0.1-1$\,TeV energy range purely by the KN effect.

\begin{figure}
\epsscale{1.0}
\plotone{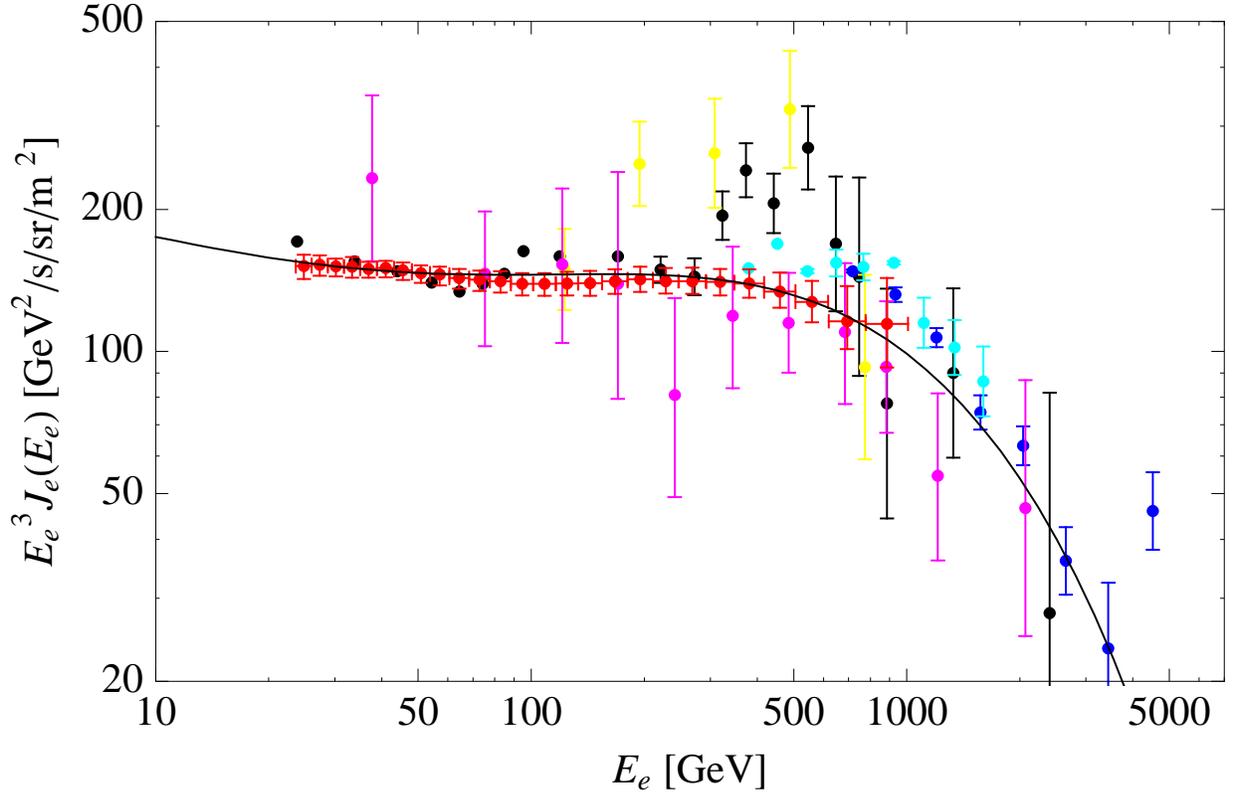}
\caption{Comparison of the observed spectra of the Galactic CR
electrons with model spectra calculated for $u_{star} = 3$\,eV\,cm$^{-3}$,
$n_{ism} = 1$\,cm$^{-3}$, $u_{dust} = 0.1$\,eV\,cm$^{-3}$, $B = 3$\,$\mu$G,
$\ell = 3$\,kpc, $s_e = 2.42$ and $E_{e,\,max} = 2.75$\,TeV. The black solid line
corresponds to the energy spectrum of the primary CR electrons calculated using
equations (4-7). The data points correspond to different measurements by ATIC 
\citep[black;][]{cha08}, PPB-BETS \citep[yellow;][]{tor08}, emulsion chambers 
\citep[magenta;][]{kob04}, HESS \citep[blue and cyan;][respectively]{aha08,aha09a}, 
and Fermi \cite[red;][]{abd09}.}
\label{CReobs}
\end{figure}

In Figure\,\ref{CReobs} we compare the observed spectra from various experiments
with one of our model calculations corresponding to a choice of model parameters
appropriate for the average (local) ISM conditions, namely $\ell = 3$\,kpc, 
$n_{ism} = 1$\,cm$^{-3}$, $B = 3$\,$\mu$G, $u_{dust} = 0.1$\,eV\,cm$^{-3}$, 
$u_{star} = 3$\,eV\,cm$^{-3}$, $E_{e,\,max} = 2.75$\,TeV, and $s_e = 2.42$. 
The data points correspond to different measurements by ATIC 
\citep[black;][]{cha08}, PPB-BETS \citep[yellow;][]{tor08}, emulsion chambers 
\citep[magenta;][]{kob04}, HESS \citep[blue and cyan;][respectively]{aha08,aha09a}, 
and Fermi \cite[red;][]{abd09}. As evident, with reasonable parameters\footnote{It
is important to note in this context that the anticipated value of the starlight 
energy density, $u_{star} = 3$\,eV\,cm$^{-3}$, even though considered here as a 
`reasonable' one, is still larger than that expected for the local ISM 
\citep{str00,por06,mos06,por08}. As such, it should be considered as an illustrative 
model assumption, for which the analyzed KN effects are already of a major importance
(see the related discussion in \S3.4 and \S4 further below).} 
\emph{and the electron injection index the same as required for the 
Galactic CR protons}, $s_e = s_p = 2.42$ (see section 3.1 below), we can 
reproduce very well the latest, and most reliable observations by Fermi and HESS.
It is also clear that the KN effect cannot account for the sharp feature claimed
by ATIC observations. Encouraged by this simple and robust explanation for the primary 
CR electron spectrum, in the next section we also explore the influence of the KN effect 
on the expected spectra of secondary \epm pairs with the goal of providing explanation 
of the rise with energy of the positron to electron ratio observed by PAMELA.

\section{Secondary and Tertiary Pairs}

In this section we address the question of the origin and spectrum of
ultrarelativistic positrons present in the CR population, which are produced as
secondaries in \epm pair production processes. We consider three different
sources of secondary \epm pairs and apply the same transport equation as above
to determine their spectra in the ISM. From these we obtain the positron to 
electron ratio and compare it to the observation by PAMELA.

\subsection{Proton-Proton Pair Production}

The first source of secondary pairs we consider is due to the interactions 
of ultrarelativistic CR ions (primarily protons) with the ambient plasma. We
assume that the Galactic sources of CRs in addition to electrons inject also
ultrarelativistic protons at a constant rate ${\dot Q}_p(E_p) \propto
E_p^{-s_p}$, which then propagate diffusively through the ISM and collide 
with cold protons. The appropriate timescale for the proton-proton
interaction is roughly independent of energy: $\tpp \simeq (c \, n_{ism}
\, \sigma_{pp})^{-1} \simeq 30\,(n_{ism}/{\rm cm^{-3}})^{-1}$\,Myr for the
cross-section $\sigma_{pp} \simeq 3.4 \times 10^{-26}$\,cm$^2$ \citep[see, 
e.g.,][]{kel06}. The diffusive escape timescale for CR protons is same as for 
electrons, namely $\tesc \simeq 100\,(E_p/{\rm GeV})^{-1/3}$\,Myr (for 
$\ell \simeq 3$\,kpc, and $n_{ism} \simeq 1$\,cm$^{-3}$). This means that 
CR protons with $E_p > 30$\,GeV are in a slow cooling regime (i.e., we are 
dealing with a thin target case), so that the ISM proton energy spectrum 
can be approximated as $n_p(E_p > 30\,{\rm GeV}) \simeq \tesc \times {\dot Q}_p(E_p) 
\propto E_p^{-s_p - 1/3}$ (for the Kolmogorov turbulence; see the discussion
in section 2.1 and below equation\,\ref{solution}). 
Keeping in mind the observed CR proton flux $J_p(E_p) \propto E_p^{-2.75}$, 
the required injection spectral index should be then $s_p  \simeq 2.42$. In
addition, in this regime protons escape with most of their energy and only a
small fraction $f$ of the carried flux goes into production of secondaries (\epm and
neutrinos arising from $\pi^\pm$ decays) and $\gamma$-rays (from $\pi^0$ decay).
In particular, one has $f \simeq \tesc /\tpp \simeq 0.3 \, (E_p/{\rm TeV})^{-1/3}$. 
Note that if the CR protons propagate through the ISM with some particular 
bulk velocity, e.g., of the order of the Alfven speed, the situation may change. 
For example, with $\tau_{dyn} \simeq \ell / v_A$ one gets $f\simeq 
\tau_{dyn}/\tpp \simeq 10$ independent of the proton energy
(for $B \simeq 3$\,$\mu$G, $\ell \simeq 3$\,kpc, and $n_{ism} \simeq 1$\,cm$^{-3}$).
In this case one would expect $n_p(E_p) \simeq \tpp \times {\dot Q}_p(E_p) \propto
E_p^{-s_p}$, requiring thus a steeper injection index of $s_p \simeq 2.75$.

Independent of which CR proton propagation model is the correct one, the
production rate of the secondary pairs will depend on the observed spectrum of
the CR protons (which we assume to be the same throughout the Galactic disk):
\beq\label{Qepm}
{\dot Q}_{e^\pm}(E_{e^\pm}) \simeq \tau_{pp}^{-1} \,\, n_p(E_p) \,\, 
f_{e^\pm}\!\!\left(E_{e^\pm}/E_p\right) \, ,
\eeq
where $f_{e^\pm}(E_{e^\pm}/E_p)$ is the number of pairs with energy $E_{e^\pm}$ 
produced by a CR proton of energy $E_p$. Detailed calculations by \citet{kel06} 
show that for $1$\,TeV\,$\lesssim E_p \lesssim 1$\,PeV, the function
$f_{e^\pm}(E_{e^\pm}/E_p)$ is strongly peaked for $E_{e^\pm}/E_p \simeq 0.07$ 
at the level $f_{e^\pm}^{max} \simeq f_{e^\pm}(0.07) \simeq 4$. As a result, 
the injection function of the secondary pairs should follow the energy spectrum 
of CR protons, namely ${\dot Q}_{e^\pm}(E_{e^\pm}) \propto E_{e^\pm}^{-s_p}$. 
Since the secondary pairs obey the same transport equation as the primary electrons, 
their spectra can be calculated as discussed in \S\,2. In particular, for high energies 
the escape term (as well as the Coulomb and bremsstrahlung energy losses)
can be ignored giving $n_{e^\pm}(E_{e^\pm} >{\rm 10 GeV}) \simeq 
\tr \times {\dot Q}_{e^\pm}(E_{e^\pm}) \propto E_{e^\pm}^{-s_p - 1}$ for
$\tr \simeq \trT$. More generally, the expected secondary pair to total electron
ratio should vary with the energy roughly as
\bar
\left.{n_{e^\pm} \over n_e}\right|_{pp} & \simeq & {\tr(E_e) \over \tpp} \,\, {4 \,
J_p( 14E_e) \over J_e(E_e)} \nonumber \\
& \xrightarrow[\hspace{7pt}\hspace{7pt}]{\rm T} & 4 \, \left({n_{ism} \over
{\rm cm}^{-3}}\right) \, \left({u_{star} \over {\rm eV \, cm^{-3}}}\right)^{-1}
\, \left({E_e \over {\rm GeV}}\right)^{-0.75} \, ,
\ear
where $J_p(E_p) \simeq 2.2 \times 10^4 \, (E_p/{\rm GeV})^{-2.75}$\,GeV$^{-1}$\,m$^{-2}$\,s$^{-1}$\,sr$^{-1}$ 
is the observed CR proton flux, and 
$J_e(E_e) \simeq 155\,(E_e/{\rm GeV})^{-3}$\,GeV$^{-1}$\,m$^{-2}$\,s$^{-1}$\,sr$^{-1}$ 
is the observed CR electron flux. The last line in the above equation assumes we are 
in the Thomson regime with $\tr \simeq \trT \propto E_{e^\pm}^{-1}$ so that this 
ratio becomes $n_{e^\pm}/n_e \propto E_e^{-0.75}$ with the particular value 
$(n_{e^\pm} /n_e)_{100\,{\rm GeV}} \simeq 0.04$ for the assumed starlight 
density of $u_{star} \simeq 3$\,eV\,cm$^{-3}$. This is in a disagreement with the 
PAMELA results indicating the \epm fraction increasing with energy up to $(n_{e^{\pm}} 
/ n_e)_{100\,{\rm GeV}} > 0.1$ \citep{adr09}. However, as discussed above, at higher 
energies we are in the KN regime, where $\tr \simeq \trKN \propto E_e^{1/2}$, which 
will give rise to a flatter energy spectrum of the secondary pairs and hence to 
$n_{e^\pm}/n_e \propto E_{e^\pm}^{0.75}$. Note also that, since $\trT \propto 
u_{star}^{-1}$, in regions of low (high) radiative field densities the expected 
\epm fraction will be higher (lower) for a given $J_p(E_p)$ and $J_e(E_e)$.

\subsection{Photo-Pair Production}

One possibility for increasing the pair fraction in the CR spetrum is to introduce
an additional, flatter spectral component consisting solely of the \epm pairs that 
outnumber the secondaries resulting from the proton-proton interactions\footnote{Note 
that such a population cannot be accompanied by the additional population of 
proton-antiproton pairs, since this would violate the observed proton-to-antiproton 
ratio \citep{mos02}.}. However, this population cannot extend up to $E_e > 1$\,TeV 
energies, since this would violate the high-energy cut-off measured in the CR electron 
spectrum by the HESS experiment \citep{aha08}. A possible source of pairs that satisfy 
these requirements may be due to photon-photon annihilation of TeV-energy 
$\gamma$-rays on starlight \citep{aha91,mas91b}. The cross-section for this process has a 
sharp peak when photon energies satisfy the condition $\varepsilon_0 \, \varepsilon_{\gamma}
= 2 \, m_e^2 c^4$. Thus, the annihilation of $\varepsilon_0 \simeq
\varepsilon_{star} \simeq 1$\,eV and $\varepsilon_{\gamma} \simeq 0.5$\,TeV $\gamma$-ray
photons will inject into the ISM a relatively narrow energy distribution of pairs at the rate
${\dot Q}_{{e^\pm},\,\gamma\gamma}(E_{e^\pm}) \propto \delta (E_{e^\pm} - m_e^2c^4/\varepsilon_{star})$.
Such a distribution cooling radiatively according to equation (\ref{simpkeq})
will produce a flat-spectrum $n_{e^\pm}(E_{e^\pm}) \propto E_{e^\pm}^{-2}$ (in the
Thomson regime, or even a flatter one in the KN regime), instead of $\propto 
E_{e^\pm}^{-3.75}$ expected for the secondaries resulting from the decay of $\pi^{\pm}$ generated
in the proton-proton interactions, as described above.

However, the question is whether there will be sufficient number of such pairs
to account for the PAMELA observations. In order to address this issue we use
the $\delta$-function approximation for the photon-photon annihilation cross 
section $\sigma_{\gamma \gamma}(\varepsilon_0, \varepsilon_\gamma) \simeq (1/3) \,
\sigma_T \, \varepsilon_{0} \, \delta[\varepsilon_{0} - (2 m_e^2 c^4/\varepsilon_{\gamma})]$ 
\citep{zdz85}, from which we can calculate the absorption coefficient $\alpha_{\gamma 
\gamma}(\varepsilon_\gamma)= \int_{m_e c^2/\varepsilon_{0}} d\varepsilon_{0} \, 
n_{0}(\varepsilon_{0}) \, \sigma_{\gamma \gamma}$. If we also approximate the 
energy density of the soft (starlight) photon field by a monoenergetic
distribution with total density $n_{star}$ and energy $\varepsilon_{star}$, namely
$n_{0}(\varepsilon_0) = n_{star}\delta(\varepsilon_0 - \varepsilon_{star})$
such that $\int d\varepsilon_0 \, u_{0}(\varepsilon_0) = u_{star}= n_{star} \, 
\varepsilon_{star}$, then the opacity becomes $\alpha_{\gamma \gamma}(\varepsilon_\gamma) 
\simeq (\sigma_T / 3) \, u_{star} \, \delta[\varepsilon_{star} - (2 m_e^2 c^4/ 
\varepsilon_{\gamma})]$. From this we can evaluate the optical depth to be
\beq\label{optdepth1}
\tau_{\gamma \gamma}(\varepsilon_\gamma) \simeq \tau_{\gamma \gamma}^0 \times
\delta\!\!\left[\varepsilon_{star} - {2 m_e^2 c^4 \over \varepsilon_{\gamma}}\right] \, ,
\eeq
where
\beq\label{optdepth2}
\tau_{\gamma \gamma}^0 \equiv {1 \over 3} \, \ell \, \sigma_T
\, u_{star} \, \varepsilon_{star}^{-1} \simeq 2 \times 10^{-3} \, \left({\ell
\over 3\,{\rm kpc}}\right) \, \left({u_{star} \over {\rm eV\,cm^{-3}}}\right) \,
\left({\varepsilon_{star} \over {\rm eV}}\right)^{-1} \, .
\eeq
Since $\tau_{\gamma \gamma}^0 \ll 1$, most of the $\gamma$-rays freely escape
the Galaxy and thus their number density per energy is $n_{\gamma}(\varepsilon_{\gamma}) 
\simeq (\ell/c) \, {\dot Q}_{\gamma}(\varepsilon_{\gamma})$, where 
${\dot Q}_{\gamma}(\varepsilon_{\gamma})$ is the $\gamma$-ray production rate discussed below.

\subsubsection{Tertiary Pairs from Hadronic Interactions}

Proton-proton interactions, in addition to producing secondary pairs, also
produce $\gamma$-rays (from $\pi^0$ decay) of similar spectrum and comparable
intensity. These $\gamma$-rays could be the source of the \epm pairs (which may
be called tertiary pairs) in the above scenario. The rate of such `hadronic' 
$\gamma$-ray production may be approximated as
\beq\label{ppgrate}
{\dot Q}_{\gamma,\,pp}(\varepsilon_{\gamma}) \simeq \tau_{pp}^{-1} \,\, n_p(E_p) \,\,
f_{\gamma}(\varepsilon_{\gamma}/E_p) \, ,
\eeq
where
$f_{\gamma}(\varepsilon_{\gamma}/E_p)$ is the number of photons with energy
$\varepsilon_{\gamma}$ produced in a single proton-proton collision involving a
CR proton with the energy $E_p$. Just as in the case of secondary pair production,
we refer to \citet{kel06}, who showed that in the range $0.1$\,TeV\,$\lesssim E_p 
\lesssim 1$\,PeV the function $f_{\gamma}(\varepsilon_{\gamma}/E_p)$ is peaked 
for $\varepsilon_{\gamma}/E_p \simeq 0.1$ at the level $f_{\gamma}^{max} \simeq 
f_{\gamma}(0.1) \simeq 6$.

The total production rate of such tertiary \epm pairs (with energies $E_e 
\simeq \varepsilon_{\gamma}/2$) may be obtained from ${\dot Q}_{{e^\pm},\,\gamma
\gamma}(E_e) \simeq \left. 4 c \, \alpha_{\gamma \gamma} \, n_{\gamma}
(\varepsilon_{\gamma})\right|_{\varepsilon_{\gamma} = 2 E_e}$ \citep{cop90}.
As before, inserting this in equation (\ref{simpkeq}) and carrying out the
integration, we get the density ratio of tertiary pairs to total electrons at 
the same energy $E_e$ in terms of the observed CR proton and electron
flux ratio,
\bar
\left.{n_{e^\pm} \over n_e}\right|_{\gamma \gamma/pp} & \simeq & 24 \,
\tau_{\gamma
\gamma}^0 \,\, {\tr(E_e) \over \tpp} \, {m_e^2 c^4 \over E_e \,
\varepsilon_{star}}\,\, {
J_p( 20 m_e^2 c^4 / \varepsilon_{star}) \over J_e(E_e)} \nonumber \\
& \xrightarrow[\hspace{7pt}\hspace{7pt}]{\rm T} & 10^{-6} \, \left({\ell \over
{\rm 3\,kpc}}\right) \, \left({n_{ism} \over {\rm cm}^{-3}}\right) \,
\left({\varepsilon_{star} \over {\rm eV}}\right)^{0.75} \, \left({E_e \over 
{\rm GeV}}\right)\, ,
\ear
where for the the bottom line we have assumed $u_{star} \simeq u_{tot}$ and used
the Thomson regime for $\tr(E_e) $.

First, we note that because $\tau_{\gamma \gamma}^0 \ll 1$ (see equation\,\ref{optdepth1}) 
the expected number of (tertiary) pairs from photo-pair process will be lower than 
that of the (secondary) pairs from proton-proton interaction. Second, because 
$\tau_{\gamma \gamma}^0 \propto u_{star}$ and in the Thomson regime $\tr \propto 
u_{star}^{-1}$, the ratio of photo-pairs to primary electron is independent of the
energy density of the soft photon field, as long as it dominates over the 
other Galactic photon fields and the magnetic field. However, more importantly, 
this ratio increases with increasing starlight energy as $n_{e^\pm}/n_e \propto 
\varepsilon_{star}^{0.75}$, and (in the Thomson regime) it increases linearly
with electron energy. Therefore, in regions of the Galaxy containing high energy 
(ultraviolet) photons and for high energy electrons the photo-pair production may 
become important and even dominant (see below). Of course, the above result again 
will be modified by  the inevitable KN effect. In this contex it should be 
emphasized that because of the flatter injection function of the tertiary pairs 
resulting from the photon-photon annihilation the KN effect should be more pronounced 
for them than for the primary electrons or the secondary pairs originating from the
decay of $\pi^{\pm}$ due to proton-proton collisions.

\subsubsection{Tertiary Pairs from Leptonic Interactions}

Yet another source of $\gamma$-rays which may annihilate on the starlight photon
field and create additional \epm population is provided by the IC emission of CR 
electrons themselves. In order to estimate the expected relevance of this process, 
we need the rate of production of $\gamma$-rays, ${\dot Q}_{\gamma,\,ic}(\varepsilon_{\gamma})$, 
which will take the place of ${\dot Q}_{\gamma,\,pp}(\varepsilon_{\gamma})$ specified in the 
previous section. The IC rate is related to the IC emissivity $j_{ic}(\varepsilon_{\gamma})$ 
as ${\dot Q}_{\gamma,\,ic}(\varepsilon_{\gamma})=4\pi j_{ic}(\varepsilon_{\gamma})/\varepsilon_{\gamma}$,
which can be obtained from the standard relation $[\varepsilon_{\gamma} j_{ic}(\varepsilon_{\gamma})] 
\simeq (1 / 4 \pi) \, [E_{e,\,ic}^2 \, n_e(E_{e,\,ic})] / \tr\!(E_{e,\,ic})$. Here $E_{e,\,ic}$ 
is the energy of electrons emitting $\gamma$-ray photons with energies $\varepsilon_{\gamma}$, 
while $\tr\!(E_{e,\,ic})$ includes only IC cooling due to soft photons of energy $\varepsilon_{star}$. 
This gives
\beq\label{icgrate}
{\dot Q}_{\gamma,\,ic}(\varepsilon_{\gamma}) = \left({E_{e,\,ic} \over \varepsilon_{\gamma}}\right)^2 \,
{n_e(E_{e,\,ic}) \over \tr\!(E_{e,\,ic})} \, ,
\eeq
which replaces the photon production rate given above in equation (\ref{ppgrate}).

Following the same procedure as above we can evaluate the density of $\gamma$-rays,
the rate of production of \epm pairs ${\dot Q}_{{e^\pm},\,\gamma\gamma}(E_{e^\pm})$,
and then the density of pairs in the ISM. In the Thompson regime $\varepsilon_{\gamma}
\simeq (4/3) \, E_{e,\,ic}^2\, \varepsilon_{star}/m_e^2c^4$ and thus $(E_{e,\,ic} / 
\varepsilon_{\gamma})^2= 3/8$ for $\varepsilon_{star} = 2 m_e^2 c^4 / \varepsilon_{\gamma}$, 
while in the KN regime $E_{e,\,ic} / \varepsilon_{\gamma} \simeq 1$. As we will see below, 
for relevant CR energies we are closer to the KN regime so we will ignore the factor $3/8$. 
We then obtain
\bar
\left.{n_{e^\pm} \over n_e}\right|_{\gamma \gamma/ic} & \simeq & 4 \, \tau_{\gamma
\gamma}^0 \,{m_e^2 c^4 \over E_e \, \varepsilon_{star}} \, {\left[J_e(E_e) / 
\tr)\right]_{E_e = 2 m_e^2 c^4/\varepsilon_{star}} \over \left[J_e(E_e)/\tr\right]} \, 
\nonumber \\ & \xrightarrow[\hspace{7pt}\hspace{7pt}]{\rm T} & 8 \times 10^{-6} \,
\left({\ell \over {\rm 3\,kpc}}\right) \, \left({u_{star} \over {\rm eV \,
cm^{-3}}}\right) \, \left({E_e \over {\rm GeV}}\right) \,
\ear
where the last line is evaluated for the Thomson regime with $\tr \propto E_e^{-1}$.
As evident, in this regime and for the observed $J_e(E_e)$ we obtain 
$(n_{e^\pm}/n_e)_{100\,{\rm GeV}} \simeq 2 \times 10^{-3}$ 
for $\ell \simeq 3$\,kpc and $u_{star} \simeq 3$\,eV\,cm$^{-3}$, independent of 
the soft photon energy. This implies that the production of TeV-energy $\gamma$-rays 
via IC emission of CR electrons --- if proceeding in the Thomson regime --- may 
dominate over the one resulting from the protons-proton interactions. On the other 
hand, the KN effect are expected to reduce the IC emissivity of ultrarelativistic 
\epm pairs within the consider photon energy range, and therefore both hadronic and 
leptonic processes may be in fact comparable. For the choice of model parameters 
appropriate for the average ISM conditions, this is not enough to account for the 
high positron fraction found in the CR spectrum.

\subsection{Energy Spectra of Secondaries and Tertiaries}

As in case of primary electrons we now carry a more accurate determination of
secondary and tertiary pairs by omitting most of the approximations used above.
The production rate of the secondary pairs from proton-proton interactions is
now obtained from
\beq
{\dot Q}_{pp}(E_e) = {1 \over \tpp} \, \int_{E_e} \, {d E_p \over E_p} \,
J_p(E_p) \, f_e(E_p, E_e) \, ,
\eeq
where we use the analytic approximation for the function $f_e(E_p, E_e)$ as 
given in \citet{kel06}, and fix $J_p(E_p) \simeq 2.2 \times 10^4 \, (E_p / 
{\rm GeV})^{-2.75}$\,GeV$^{-1}$\,m$^{-2}$\,s$^{-1}$\,sr$^{-1}$.
For the production rates of (tertiary) pairs generated from  annihilation of
high-energy $\gamma$-rays with density $n_\gamma(\varepsilon_\gamma)$ by the
soft Galactic photon fields, we write analogously
\beq
{\dot Q}_{\gamma \gamma}(E_{e^\pm}) = {4 \over 3} \, \sigma_T \, c \,
\left.u_{rad}(\varepsilon)\right|_{\varepsilon = m_e^2 c^4 / E_{e^\pm}}
\left.n_{\gamma}(\varepsilon_{\gamma})\right|_{\varepsilon_{\gamma} = 2
E_{e^\pm}} \, ,
\eeq
where again we have used the delta function approximation for the photon-photon
annihilation cross section as before. For the expected small optical depth of 
photon-photon annihilation ($\tau_{\gamma \gamma}^0 \ll 1$, see above), the
spectrum of $\gamma$-rays resulting from the proton-proton interactions is given
by
\beq
n_{\gamma/pp}(\varepsilon_{\gamma}) = {\ell \over c \, \tpp} \,  \,
\int_{\varepsilon_{\gamma}} \, {d E_p \over E_p} \, J_p(E_p) \, f_{\gamma}(E_p,
\varepsilon_{\gamma}) \, ,
\eeq
with the function $f_{\gamma}(E_p, \varepsilon_{\gamma})$ denoting the number of
photons with energy $\varepsilon_{\gamma}$ produced in a single collision
involving ultrarelativistic proton with energy $E_p$. Again, here we take the
analytical approximation for $f_{\gamma}(E_p, \varepsilon_{\gamma})$ as given in
\citet{kel06}, noting that for $E_p \simeq 0.1$\,TeV\,$- 1$\,PeV this may be further
approximated by a simple function
\beq
f_{\gamma}(x) \simeq 2.5 \, x^{-1} \, \exp\left[-9 \, x^{0.83}\right] \, ,
\eeq
with $x \equiv \varepsilon_{\gamma}/E_p$ \citep[see in this context][]{hil05}.

Finally, for the case of $\gamma$-rays resulting from the IC emission of
ultrarelativistic CR leptons, we calculate the appropriate photon energy
spectrum as
\beq
n_{\gamma/ic}(\varepsilon_{\gamma})  = {4 \pi \ell \over c \varepsilon_{\gamma}}
 \, j_{ic}(\varepsilon_{\gamma}) \, ,
\eeq
where the IC emissivity $j_{ic}(\varepsilon_{\gamma})$ is related to the observed
electron flux using the standard IC formulae with the KN effect included \citep{blu70}. 
We fix this flux as $J_e(E_e) \simeq 155 \, (E_e/{\rm GeV})^{-3}$\,GeV$^{-1}$\,m$^{-2}$\,s$^{-1}$\,sr$^{-1}$ 
(and cutting-off exponentially at $E_e = 2$\,TeV). Inserting then the resulting \epm pair 
production rates in equation (\ref{simpkeq}), we obtain the individual and 
total \epm pair fluxes $J_{e^\pm}^{tot}(E_e) = J_{e^\pm}^{pp}(E_e)+
J_{e^\pm}^{\gamma \gamma/pp}(E_e) + J_{e^\pm}^{\gamma \gamma/ic}(E_e)$ for all
three mechanisms discussed above and for the same seven parameters $s_e$, $E_{e,\,max}$,
$u_{dust}$, $u_{star}$, $n_{ism}$, $\ell$ and $B$ used in calculation of the 
primary electron spectra.

\begin{figure}
\epsscale{0.5}
\plotone{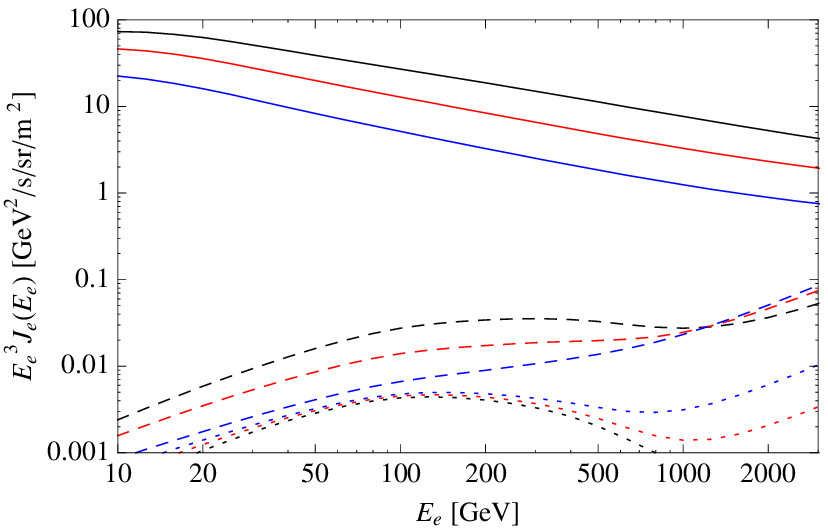}
\plotone{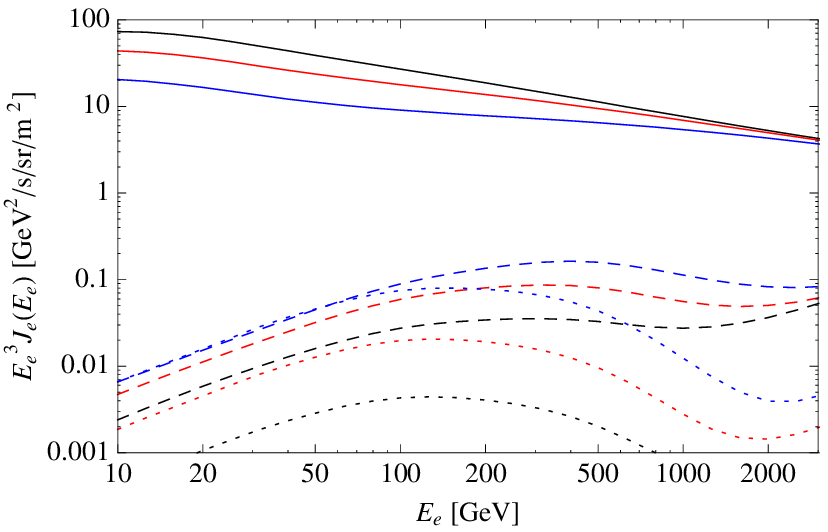}
\plotone{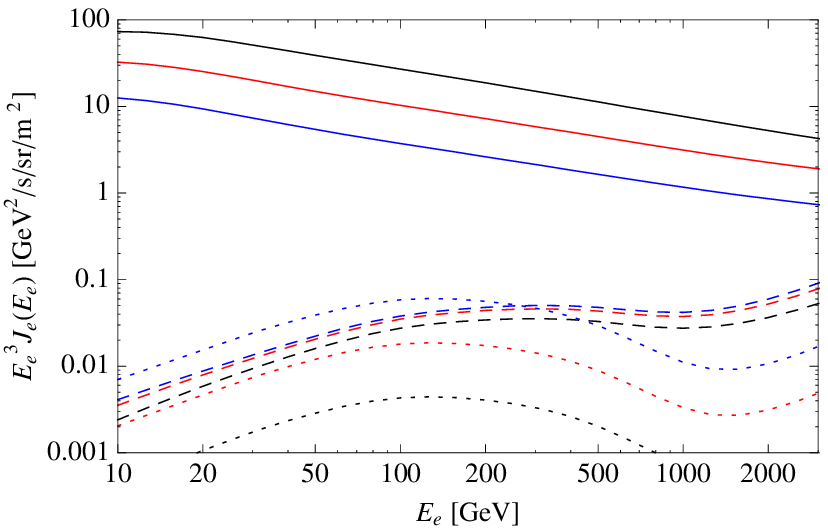}
\caption{The energy spectra of secondary leptons produced in proton-proton 
collisions (solid lines), and of tertiary pairs produced via absorption of 
high-energy $\gamma$-rays generated in either hadronic or leptonic processes 
(dashed and dotted lines, respectively), for the same set of the model free 
parameters as given in Figure\,\ref{lossrate} (black, red, and blue curves 
on different panels), except for the single value of the magnetic field 
$B= 1$\,$\mu$G.}
\label{pairs}
\end{figure}

Figure\,\ref{pairs} shows the energy spectra of secondary leptons produced in
proton-proton collisions, $J_{e^\pm}^{pp}(E_e)$ (solid lines), and of tertiary pairs 
produced via absorption of high-energy $\gamma$-rays generated in either hadronic 
or leptonic processes (dashed and dotted lines, respectively), $J_{e^\pm}^{\gamma 
\gamma/pp}(E_e)$ and $J_{e^\pm}^{\gamma \gamma/ic}(E_e)$, for the same set of parameters 
used in Figure\,\ref{lossrate} with same colors, except for the magnetic field set at 
$B= 1$\,$\mu$G. As shown, the cases with a weak starlight but strong dust emission 
are quantitatively similar to the cases when the dust and starlight energy densities 
are comparable. Only in the cases when the ratio $u_{star}/u_{dust}$ is high,
the KN effect flattens the energy distribution of secondary leptons resulting from 
proton-proton interactions significantly, and only at high ($E_e > 100$\,GeV) 
energies. In all the above cases, however, the
direct \epm pair production in the proton-proton collisions is the dominant
source of the positrons, and the contribution of the other two processes (i.e., 
of the tertiary pairs) to the positron flux is less than $1\%$ except at high 
energies where it could reach $10\%$. The contribution of secondary and tertiary 
electrons to the observed electron spectrum is even less, being on the order of 
$<10\%$ and $\lesssim 0.1\%$ for proton-proton and photo-pair processes, respectively.

\begin{figure}
\epsscale{1.0}
\plotone{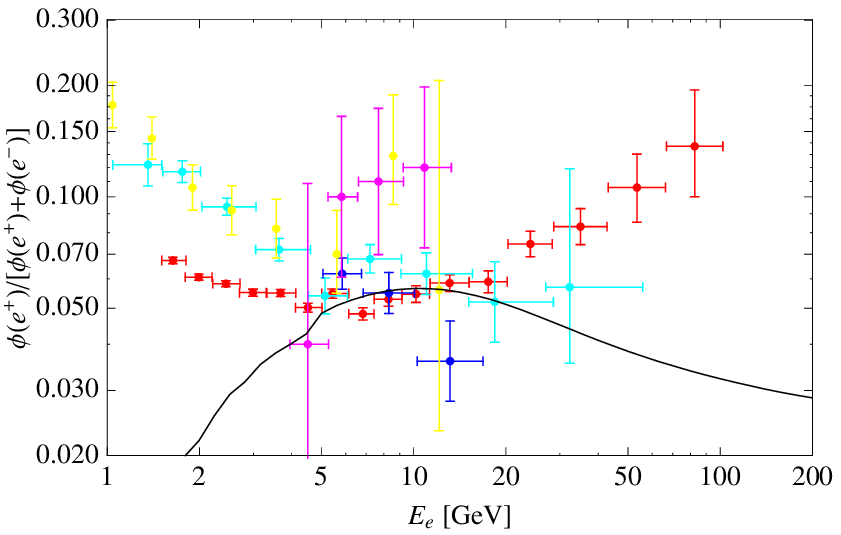}
\caption{
The positron-to-electron ratio from different measurements by PAMELA 
\citep[red symbols;][]{adr09}, HEAT \citep[cyan and blue symbols;][]{bar97,bea04}, 
CAPRICE \citep[yellow symbols;][]{boe00}, and imaging calorimeter \citep[magenta 
symbols;][]{gol94}, compared with the model result (line) for the same model 
parameters as considered in Figure\,\ref{CReobs}, namely $u_{star} = 3$\,eV\,cm$^{-3}$,
$n_{ism} = 1$\,cm$^{-3}$, $u_{dust} = 0.1$\,eV\,cm$^{-3}$, $B = 3$\,$\mu$G,
$\ell = 3$\,kpc, $s_e = 2.42$ and $E_{e,\,max} = 2.75$\,TeV.}
\label{posit-fract}
\end{figure}

The above result is illustrated in Figure\,\ref{posit-fract}, where we compare with 
different measurements the computed ratio of (both secondary and tertiary) 
positron and electron fluxes, normally denoted in the literature as $\phi(e^+)$ 
and $\phi(e^-)$, respectively,
\beq
{\phi(e^+) \over \phi(e^+) + \phi(e^-)} \equiv {1 \over 2 \, \left[J_e(E_e) /
J_{e^\pm}^{tot}(E_e)\right] + 1} \,
\eeq
for the same model parameters as considered in Figure\,\ref{CReobs} (namely 
$u_{star} = 3$\,eV\,cm$^{-3}$, $n_{ism} = 1$\,cm$^{-3}$, $u_{dust} = 0.1$\,eV\,cm$^{-3}$, 
$B = 3$\,$\mu$G, $\ell = 3$\,kpc, $s_e = 2.42$ and $E_{e,\,max} = 2.75$\,TeV). Here
red symbols denote the PAMELA data \citep{adr09}, cyan and blue symbols the HEAT 
data \citep{bar97,bea04}, yellow symbols the CAPRICE data \citep{boe00}, and the
magenta ones the measurements with the imaging calorimeter \citep{gol94}. As evident, due to the KN 
effects and inclusion of tertiary pairs the \epm fraction decreases only by a factor of 
$2$ between $E_e \simeq 10$\,GeV and $200$\,GeV. Even though this is a much less
rapid decrease than typically expected \citep[see][]{mos98}, the PAMELA results 
in the high energy ($E_e > 20$\,GeV) range cannot be reproduced with our conservative 
choice of model parameters.

\subsection{High Positron Fraction}

\begin{figure}
\epsscale{0.75}
\plotone{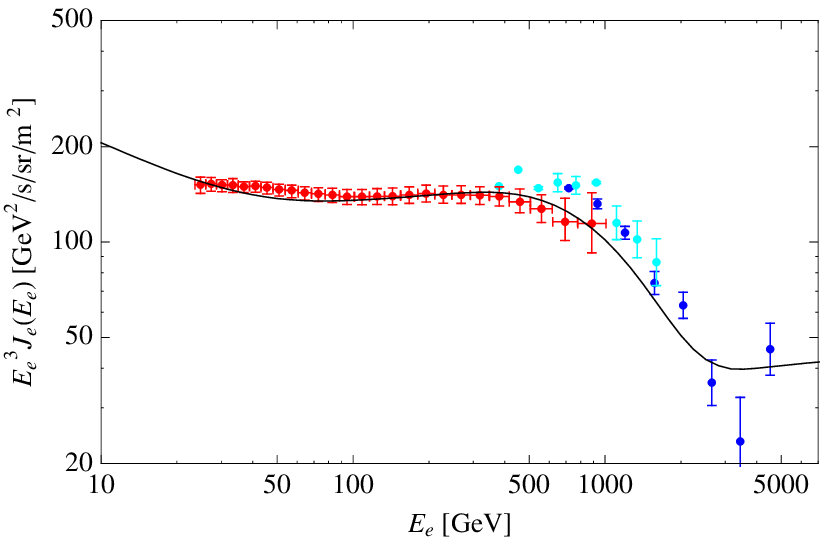}
\plotone{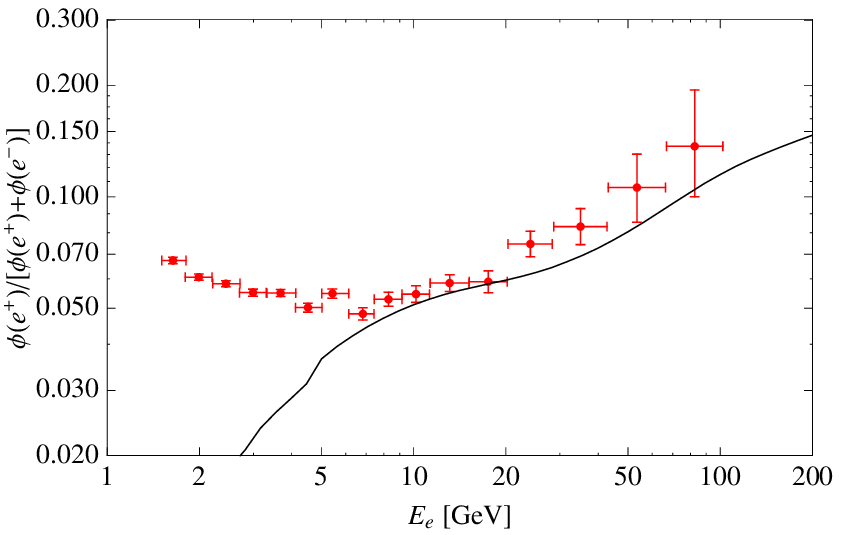}
\caption{{\bf Top panel:} The total energy spectrum of cosmic ray leptons calculated for 
$u_{star} = 300$\,eV\,cm$^{-3}$, $n_{ism} = 80$\,cm$^{-3}$, $u_{dust} = 0.1$\,eV\,cm$^{-3}$, 
$B = 3$\,$\mu$G, $\ell = 3$\,kpc, $s_{e,\,inj} = 2.65$ and $E_{e,\,max} = 1.55$\,TeV
(assuming super-exponential cut-off in the electron injection function). The data points 
correspond to the measurements by HESS (blue and cyan symbols) and Fermi (red symbols). 
{\bf Bottom panel:} The resulting positron-to-electron ratio compared with the measurements by 
PAMELA (red symbols).}
\label{PAMELA}
\end{figure}

The above results show that it would be rather difficult to increase the fraction of
secondary \epm pairs just by the photo-pair processes for the average ISM conditions. 
One needs different conditions for the production of the relatively high positron-to-electron 
fraction in the CR spectrum which also increases with energy, as claimed by the 
PAMELA experiment. In our model, those would require increasing 
the energy density of the starlight emission up to $u_{star} \sim 300$\,eV\,cm$^{-3}$, 
and of the ISM number density up to $n_{ism} \sim 80$\,cm$^{-3}$, keeping at the 
same time relatively low level of dust emission ($u_{dust} \sim 0.1$\,eV\,cm$^{-3}$) 
and magnetic field strength ($B \leq 10$\,$\mu$G). Figure\,\ref{PAMELA} (bottom panel) 
shows the \epm fraction expected for such a choice of model parameters, which agrees with the 
PAMELA data within the energy range not affected by the solar modulation. The corresponding
total electron spectrum is compared with the Fermi and HESS data in the top panel of
Figure\,\ref{PAMELA}. As evident we get again a very good agreement but now we need an
even steeper injection spectrum of the primary electrons ($s_e \simeq 2.65$).

The set of model parameters considered in Figure\,\ref{PAMELA} should be regarded 
as illustrative one only, not necessarily being justified for the local ISM. We note, 
however, that it corresponds to the optical depth for annihilation of Galactic $\gamma$-rays 
formally less than (though close to) unity (see equation\,\ref{optdepth2}), and to a small 
($<0.1$) ratio of number densities of $\gamma$-ray photons and CR electrons with the same 
energy $\varepsilon$, as required. In fact, we have
\bar
{n_{\gamma}(\varepsilon) \over n_e(\varepsilon)} \simeq {n_{\gamma,\,pp}(\varepsilon) 
\over n_e(\varepsilon)} \simeq {6 \, \ell \over c \, \tpp} \, {J_p(10\,\varepsilon) \over 
J_e(\varepsilon)} \sim 5 \times 10^{-4} \, \left({n_{ism} \over {\rm cm^{-3}}}\right) \, 
\left({\varepsilon \over {\rm GeV}}\right)^{0.25}
\ear
(see the discussion in section 3.2 above).

The invoked increased level of the starlight energy density and of the gas
number density could be more appropriate around supernova remnants where the 
injection of the Galactic CRs is taking place. Hence, the results of our analysis may
indicate that ultrarelativistic particles generated in the Galaxy undergo most of 
their interactions near their sources, but propagate much more freely from these regions to 
the Earth \citep[see in this context recent discussion in][]{hig09,cow09}. 
In fact it may be sufficient if high-energy positrons, but not necessarily 
electrons, are trapped in the regions characterized 
by the enhanced photon and gas densities. There may be even physical justification 
for such a situation. For example, CR protons streaming along large-scale magnetic field in 
the far upstream of supernova shocks with super-Alfv\'enic speed may excite resonant 
Alfv\'en waves in a form of coherent circularly-polarized cyclotron radiation 
\citep{ler67,kul69,ces80}. Due to the particular helicity of the generated waves, 
they will interact with positrons of gyroradii comparable to their wavelenghts (i.e., 
to gyroradii of CR protons generating the turbulence), but not with the electrons. 
As a result, the electrons will propagate much more freely along the Galactic magnetic 
field to the Earth, experiencing the `average' ISM conditions\footnote{Note that the 
returning current will be assured by the ambient plasma, and would involve sub-thermal 
bulk velocities of ISM particles due to the expected high number density of 
ISM within the Galactic disk.}. The results presented in Figure\,\ref{CReobs} regarding 
the observed CR electron spectrum would then be appropriate. CR positrons, 
on the other hand, will undergo enhanced scattering in vicinities of their sources 
resulting in their increased fraction in the observed CR spectrum around $100$\,GeV
energies, as presented in Figure\,\ref{PAMELA} (bottom panel). A quantitative description 
of such a possibility would require different treatment of the positron and electron 
transport within the Galaxy. This is beyond the scope of this paper. The point is, 
however, that the efficient trapping of TeV-energy CRs in vicinities of supernova remnants, 
either charge-dependent or not, may justify the high values for the starlight and gas 
densities invoked to explain the PAMELA data in a framework of our model.

Let us mention in this context that in the local environments of SNRs
additional processes may operate leading to an increase in the CR positron-to-electron 
fraction. These include enhanced interactions of freshly accelerated CR protons 
with an intense high-energy photon field of young remnants, generating thus additional 
secondary \epm pairs via the photo-mezon production process \citep{hu09}, or the 
direct acceleration of secondary pairs injected into the immediacy of SNR shocks 
via $pp$ collisions \citep{bla09}.

Yet one more process which may be relevant in the discussed context is the creation of
\epm pairs by photons in the electromagnetic field of ultrarelativistic electrons,
referred in the literature as a `triplet pair production' \citep[TPP; see][]{mas86,mas91a,der91}.
This process occurs when the energy of the incident photon in the electron rest frame exceeds 4 times
the rest energy of the electron, $\varepsilon' > \varepsilon_{cr} \equiv 4 m_e c^2$. For the 
starlight parameters considered in this paper, namely $\varepsilon \simeq 1$\,eV, 
this criterium is marginally fulfilled only in the `head-on' interactions with the 
highest energy electrons, $E_e \simeq 1$\,TeV, since only in such a case 
$\varepsilon' \simeq 2 \, \varepsilon \, E_e / m_e c^2 \sim 2 \, \varepsilon_{cr}$.
For all the other angles between the direction of an interacting electron and
starlight photon, and for all the lower-energy electrons, we have obviously
$\varepsilon' \ll \varepsilon_{cr}$. Nevertheless, the TPP may be of a primary
importance if the soft photon energies are higher than anticipated here, say
$\varepsilon \simeq 10$\,eV. Then the head-on collisions of such UV photons with the
TeV-energy electrons will produce effectively \epm pairs with energies
$E_{e^\pm} \sim 0.5 \, (E_e / \varepsilon)^{1/2} \, m_e c^2 \sim 0.1$\,TeV \citep[see][]{der91}, 
i.e. exactly within the energy range of the PAMELA excess. Note that the energy losses 
of thus produced pairs should be dominated by the IC scattering deep in the KN regime, and 
hence the spectral pile-ups discussed in this paper will flatten additionally the injected positron 
spectrum around $E_e \sim 10-100$\,GeV energies. That is because the TPP cross-section, 
$\sigma_{TPP} \sim \alpha_{fs} \times \sigma_T$, exceeds the IC cross section (due to 
the KN supression of the latter one) only for $\varepsilon' > 300 \, m_e c^2$, while the TPP 
cooling rate exceeds the IC cooling rate only for $\varepsilon' > 10^5 \, m_e c^3$ 
\citep{mas91a,der91}. As a result, if the sources of Galactic CRs are associated
with an intense UV photon field, the most recent PAMELA results may be possibly explained
with much less extreme ISM parameters than discussed in this section.

\section{Summary and Discussion}

In this paper we show that the observed excesses in the energy distribution of the 
Galactic CR electrons around energies $E_e \sim 0.1-1$\,TeV may be easily re-produced 
without invoking any unusual source of ultrarelativistic electrons (or \epm pairs), 
such as dark matter annihilation/decay or some nearby astrophysical object (e.g. a 
pulsar), other than the general diffuse Galactic components of CR electrons and 
protons injected by supernova remnants. The model presented here assumes an injected 
spectrum of electrons (power-law with index $s_e$) and evaluates their observed
energy distribution based on a simple and most commonly invoked kinetic equation 
describing the propagation of CR electrons in the ISM. The main process affecting this 
outcome is the cooling of the injected electrons by their interaction with the 
ISM photons (via IC scattering). The interactions of electrons with ISM turbulence produces
negligible re-acceleration and determines their escape time. The escape timescale
also turns out to be somewhat longer than the cooling time in the relevant range of 
electron energies. The new physical effect that is the source of the
observed excess is the Klein-Nishina suppression of the IC cooling rate, which becomes 
important right around TeV energies. With a very reasonable choice of the model 
parameters characterizing the local interstellar medium ($u_{star} \sim 3$\,eV\,cm$^{-3}$, 
$u_{dust} \sim u_{cmb} \sim 0.3$\,eV\,cm$^{-3}$, $B \sim 3$\,$\mu$G, and $n_{ism} 
\sim 1$\,cm$^{-3}$) we can reproduce the most recent, and perhaps the most reliable 
observations by Fermi and HESS, but not the sharp feature claimed by ATIC. 
Interestingly, in our model the injection spectral index of CR electrons becomes 
comparable to, or perhaps equal to that of CR protons, namely $s_e \simeq s_p \simeq 2.4$.

The Klein-Nishina effect will also affect the propagation of the secondary \epm pairs 
and can produce deviations from a power-law in the observed spectra of such pairs.
In particular, it can affect the positron-to-electron ratio. 
We have explored this possibility by considering two mechanisms for
production of \epm pairs (and therefore positrons). The first is production of
pairs due to the decay of $\pi^\pm$'s generated by interaction of CR nuclei with
ambient protons. The second source discussed here is the pair production due to 
annihilation of diffuse Galactic $\gamma$-rays interacting with the starlight 
photon field. We consider two sources of the Galactic $\gamma$-rays. The first is
related to the decay of $\pi^0$'s also produced in proton-proton interactions and the
second is due to the IC scattering of primary CR electrons by the diffuse 
Galactic photon fields. We show that indeed there will be deviations from a simple 
power-law in the spectra of thus created \epm pairs (as well as in the positron-to-electron
flux ratio), similar to the observed one. However, the relatively high observed
positron fraction that increases quite steeply with energy, as observed by PAMELA, 
cannot be explained by the conservative set of the model parameters used above, 
which corresponds to the average values expected in the Galactic disk. We can however 
reproduce the PAMELA result by increasing the energy density of the starlight photon 
field and of the ISM number density up to the levels $u_{star} \sim 300$\,eV\,cm$^{-3}$ 
and $n_{ism} \sim 80$\,cm$^{-3}$. With these new values we can also fit the Fermi 
and HESS data, though with somewhat steeper injected spectrum of the primary electrons
than required before ($s_e \sim 2.65$). 

The required increased level of the starlight energy density and of the gas
number density may be regarded as unlikely for the local interstellar medium.
However, such a choice of the model parameters could be more appropriate around
supernova remnants where the injection of the Galactic CRs is taking place. A 
possible solution to this problem may be that CRs undergo most of their 
interactions near their sources, being efficiently trapped thereby by self-generated
CR-driven turbulence. Interestingly, such a trapping may be charge-dependent, affecting
positrons more than the electrons. A possible cause of this could be if the dominant CRs, 
namely protons, generate Alfv\'en waves of a particular helicity which scatter and 
therefore trap positrons more efficiently than electrons in the regions characterized 
by the enhanced photon and gas densities. Alternatively, higher than considered here 
energies of photons associated with CR sources may reduce significantly 
the invoked `extreme' values of the model parameters, due to even more severe KN effects
and additional (triplet) pair production processes expected to occur in an intense UV 
radiation field.

We note in this context that the qualitatively similar effects to the ones
analyzed here for the case of our Galaxy have been discussed previously for the
case of the host galaxy of nearby radio source Centaurus~A by \citet{sta06}. The
theoretically predicted isotropic, galactic-scale halo of ultrarelativistic
\epm pairs thereby (with the energy distribution shaped by the KN and
$\gamma$-ray annihilation processes), and in particular the resulting TeV
emission, has been possibly already detected by the HESS instrument
\citep{aha09b}.

\acknowledgments

We are grateful to Igor V. Moskalenko, Troy A. Porter, and Micha{\l} Ostrowski 
for helpful discussions and their valuable comments to the paper. We also
acknowledge Elliott D. Bloom for pointing out the importance of the
triplet pair production process. Finally, we thank the anonymous referee
for her/his valuable remarks on the manuscript.
\L S was supported by the Polish Ministry of Science and
Higher Education through the project N N203 380336, and also by the Scandinavian
NORDITA program on `Physics of Relativistic Flows'.

\end{document}